\documentclass[trackchanges]{aastex701}

\usepackage{graphicx}
\usepackage{amsmath}
\usepackage{siunitx}
\usepackage{subcaption}
\usepackage{float}
\begin{document}

\title{CSST-PSFNet: A Point Spread Function Reconstruction Model for the CSST Based on Deep Learning}

\correspondingauthor{Peng Wei, Chao Liu}
\email{weipeng01@nao.cas.cn, liuchao@nao.cas.cn}

\author[0009-0003-4610-6142]{Peipei Wang}
\affiliation{National Astronomical Observatories, Chinese Academy of Sciences, Beijing 100101, People's Republic of China}
\affiliation{School of Astronomy and Space Science, University of Chinese Academy of Sciences, Beijing 100049, People's Republic of China}
\email{ppwang@bao.ac.cn}

\author[0000-0003-2477-6092]{Peng Wei}
\affiliation{National Astronomical Observatories, Chinese Academy of Sciences, Beijing 100101, People's Republic of China}
\affiliation{School of Astronomy and Space Science, University of Chinese Academy of Sciences, Beijing 100049, People's Republic of China}
\email{weipeng01@nao.cas.cn}

\author[0000-0002-1802-6917]{Chao Liu}
\affiliation{National Astronomical Observatories, Chinese Academy of Sciences, Beijing 100101, People's Republic of China}
\affiliation{School of Astronomy and Space Science, University of Chinese Academy of Sciences, Beijing 100049, People's Republic of China}
\affiliation{Institute of Frontiers in Astronomy and Astrophysics, Beijing Normal University, Beijing, 100875, China}
\affiliation{Zhejiang Lab, Hangzhou, Zhejiang, 311121, China}
\email{liuchao@nao.cas.cn}

\author[0000-0001-6767-2395]{Rui Wang}
\affiliation{National Astronomical Observatories, Chinese Academy of Sciences, Beijing 100101, People's Republic of China}
\affiliation{School of Astronomy and Space Science, University of Chinese Academy of Sciences, Beijing 100049, People's Republic of China}
\email{wangrui@nao.cas.cn}

\author[0000-0002-9847-7805]{Feng Wang}
\affil{Center For Astrophysics, Guangzhou University, Guangzhou, Guangdong 510006, People's Republic of China}
\affil{Great Bay Center, National Astronomical Data Center, Guangzhou, Guangdong 510006, People's Republic of China}
\email{fengwnag@gzhu.edu.cn}

\author[0000-0001-7314-4169]{Xin Zhang}
\affiliation{National Astronomical Observatories, Chinese Academy of Sciences, Beijing 100101, People's Republic of China}
\affiliation{School of Astronomy and Space Science, University of Chinese Academy of Sciences, Beijing 100049, People's Republic of China}
\email{zhangx@bao.ac.cn}

\begin{abstract}
This paper presents \texttt{CSST-PSFNet}, a deep learning method for high-fidelity point spread function (PSF) reconstruction developed for the Chinese Space Station Survey Telescope (CSST). The model integrates a residual neural network, a lightweight Transformer architecture, and a variational latent representation to address key challenges in CSST imaging, including severe PSF undersampling, inter-band variability, and smooth spatial variation across the focal plane.
Trained and validated on high-resolution star–PSF pairs generated by the CSST Main Survey Simulator, \texttt{CSST-PSFNet} achieves improved pixel-level accuracy and more precise recovery of shape parameters relevant to weak lensing compared to widely used \texttt{PSFEx}.
On both the standard test dataset and a blurred dataset representing the upper bound of expected on-orbit PSF degradation, the model achieves a size residual precision below 0.005 and an ellipticity residual precision below 0.002.
A weak-label adaptation experiment further shows that the model can recover \texttt{PSFEx}-level performance when the true PSF is unknown, demonstrating robustness in controlled degradation scenarios and weak-label adaptation experiments.
These results indicate that \texttt{CSST-PSFNet} provides a flexible and extensible framework for future on-orbit PSF calibration in large-scale CSST surveys, with potential applications in weak-lensing cosmology and precision astrophysical measurements.
\end{abstract}


\section{Introduction} 
Over the past few decades, space-based observatories have revolutionized our exploration of the cosmos, opening an unprecedented window into the Universe that ground-based telescopes cannot fully replicate. By operating beyond Earth's atmosphere, these instruments eliminate distortions caused by atmospheric turbulence, enabling stable, diffraction-limited imaging with unparalleled clarity and precision. Iconic space telescopes such as \textit{the Euclid Telescope} (Euclid)\citep{Euclid_2025}, \textit{the James Webb Space Telescope} (JWST)\citep{Treu_2022, Paris_2023, Bergamini_2023}, and \textit{the Nancy Grace Roman Space Telescope} (Roman)\citep{Spergel_2015, Akeson_2019} have played a pivotal role in the advancement of astronomical research, delivering groundbreaking insights into galaxy formation and evolution, the growth of cosmic structures, the nature of dark energy and dark matter, and the origins of the Universe itself. These advances underscore the critical importance of space telescope development in pushing the boundaries of human knowledge and fostering interdisciplinary collaborations.

The \textit{Chinese Space Station Survey Telescope} (CSST) is a next-generation space survey mission in China designed to perform a ten-year optical and near-ultraviolet imaging and slitless spectroscopic survey while co-orbiting with the Chinese Space Station \citep{csstcollaboration2025introductionchinesespacestation, Zhan_2021}. With a 2-meter off-axis three-mirror anastigmat (TMA) optical system and an instantaneous field of view of approximately $1.1\,{\rm deg}^2$, CSST aims to conduct a wide and deep survey of $\sim 17{,}500\,{\rm deg}^2$, enabling precision cosmology through weak gravitational lensing, galaxy clustering, and photometric redshift measurements. Its focal plane consists of 30 detectors totaling 2.6 billion pixels, with an angular scale of \SI{0.074}{\arcsecond} per pixel and a nominal image quality of $R_{\mathrm{EE80}}<\SI{0.15}{\arcsecond}$, meaning that 80\% of the point-source energy is enclosed within roughly two pixels. Eighteen CCDs are dedicated to seven broad-band photometric channels ($NUV$, $u$, $g$, $r$, $i$, $z$, $y$), while twelve CCDs serve three slitless spectroscopic channels ($GU$, $GV$, $GI$), covering 255--1000 nm.

The configuration of CSST enables high-resolution imaging of billions of galaxies and stars, but it also presents significant challenges for image calibration and point spread function (PSF) modeling. First, the detectors are severely undersampled, with each stellar image spanning only about 1.5--2 pixels across its core (compared to finer sampling in missions like HST), which limits the direct recovery of high-frequency structures. Second, the large multi-CCD focal plane, covering seven photometric bands, exhibits noticeable differences between the CCD PSF due to optical alignment and wavelength-dependent effects. Third, even within a single CCD, the PSF varies smoothly but significantly with position due to residual optical aberrations and field-dependent distortions.

Achieving precise modeling of the Point Spread Function (PSF) is crucial to maximize the scientific capabilities of the CSST, given the complex instrumental challenges discussed earlier. 
Accurate PSF characterization is essential for the CSST's ability to make milliarcsecond-level astrometric measurements. Any shift in the PSF-induced centroids can bias the estimates of proper motion and parallax \citep{Nie_2025}.
Moreover, precise PSF modeling is indispensable for key cosmological objectives of the CSST, such as studying weak gravitational lensing and large-scale structures based on photometric redshifts \citep{Gong_2019, Gong_2025, Cao_2018}. 
During weak lensing analyses, gravitational shear signals cause galaxy shape distortions at the percent level \citep{Bartelmann_2001}. These distortions can be comparable to or even lower than the artificial ellipticities introduced by the instrumental PSF \citep{Kaiser_1995}. 
Consequently, inaccuracies in PSF estimation can introduce biases, such as underestimating shear signals or miscalibrating flux measurements, thereby compromising the telescope's ability to probe dark energy parameters with a targeted precision of better than 1\% or to map the large-scale structure of the Universe across cosmic epochs. 
Furthermore, with the CSST's ambitious survey objectives—imaging billions of galaxies at sub-arcsecond resolution over a decade and processing a 2.6 billion-pixel focal plane across multiple bands—the demand for efficient, scalable PSF reconstruction methods becomes paramount, especially in the era of big data astronomy.

PSF modeling is a long-standing challenge in modern optical telescope data processing, and significant progress has been made. PSF modeling techniques have evolved from analytical models based on physical principles to data-driven and deep learning methods. Early parametric models, such as \texttt{TinyTim} \citep{Krist1993} and the effective formalism of PSF (\texttt{ePSF}) \citep{Anderson_2000, Anderson2006}, provided interpretable optical representations but lacked the flexibility to describe complex field-dependent variations. Data-driven methods, such as Resolved Component Analysis (RCA) \citep{Ngolè_2016} and \texttt{PSFEx} \citep{Bertin_2011}, enhance adaptability by directly learning PSF variations from stellar images. In contrast, hybrid models such as WaveDiff \citep{Liaudat_2023} integrate optical priors into differentiable frameworks for end-to-end calibration. In recent years, deep learning has emerged as a powerful alternative for PSF reconstruction, offering strong representational capacity for non-linear and high-dimensional dependencies. Convolutional and encoder--decoder networks can directly infer high-fidelity PSFs from undersampled stellar images, outperforming classical interpolation techniques in both precision and generalization \citep{Jia_2020, Jia_2021, Lanusse_2021}. Generative and variational architectures further enable unsupervised learning of PSF manifolds and physical consistency by coupling latent representations with optical constraints \citep{Carney_2024, Ni_2024, Bai_2025}.

Despite substantial advances in PSF modeling techniques, ranging from traditional parametric approaches to sophisticated deep learning frameworks, directly applying these existing methods to the CSST presents significant challenges. CSST's unique instrumental characteristics, including severe undersampling where stellar images span only 1.5--2 pixels, pronounced inter- and intra-CCD PSF variations due to multi-band observations, and optical aberrations, demand tailored solutions that account for these specifics. 
Conventional models often assume well-sampled data or simpler optical systems, leading to suboptimal performance in capturing the non-analytic, wavelength-dependent, and field-dependent PSF structures inherent to CSST. 
Moreover, these methods typically depend on image stacking and polynomial fitting across a large number of stellar samples to achieve adequate precision, making them inefficient and data-hungry. Together, these challenges underscore the need for more flexible and data-driven approaches. Deep learning offers a powerful framework for this purpose, capable of generalizing spatial and instrumental correlations and achieving high reconstruction accuracy even with limited stellar samples.

In this work, we present \texttt{CSST-PSFNet}, a conditioned deep generative model for high-fidelity PSF reconstruction. The paper is organized as follows.
Section~\ref{sec: data_simulation} describes the simulated datasets and preprocessing procedures.
Section~\ref{sec: methodology} introduces the model architecture, training setup, and evaluation metrics.
Section~\ref{sec: experiments} compares \texttt{CSST-PSFNet} with \texttt{PSFEx} on both standard and in-orbit degradation test sets.
Section~\ref{sec: discussion} discusses potential strategies for adapting the model when the true orbit PSF is unknown.
Finally, Section~\ref{sec: conclusion} concludes the paper.

\section{Data Preparation}
\label{sec: data_simulation}
All simulated datasets are generated using the CSST Main Survey Simulator (\texttt{csst\_msc\_sim})\footnote{\url{https://csst-tb.bao.ac.cn/code/csst-sims/csst_msc_sim}.}, an end-to-end image simulation framework built upon \texttt{GalSim} \citep{2015A&C....10..121R} that incorporates the latest CSST optical design, detector characteristics, and survey configuration \citep{wei_2025}.
The simulator renders multi-band stellar images by convolving intrinsic source profiles with wavelength-dependent PSFs while modeling instrumental effects such as field distortion, pixel integration, and charge diffusion. In csst\_msc\_sim, the optical PSF library is derived from the opto-mechanical-thermal (OMT) model of \citet{Ban_2022, Ban_2025}, covering both static (e.g., fabrication, alignment, surface figure, CCD planarity) and dynamic aberrations (e.g., thermal drift, vibration, pointing jitter).
For each of the 18 photometric CCDs, PSFs are precomputed on a $30\times30$ spatial grid across four wavelength sub-bands per filter, interpolated to arbitrary coordinates, and combined via throughput weighting.  
Each rendered stellar image is linked to its PSF through a unique \texttt{obj\_id} in the output catalog. In all datasets, the PSFs generated by the CSST simulator are used exclusively as ground-truth supervision in the loss function and are not provided to the network as input features.

NGC 2298 has been designated as one of the primary on-orbit calibration targets for CSST and will be observed using a multi-exposure dither strategy to ensure complete focal plane coverage.  
\texttt{DATASET 1} is constructed from simulated observations of this field, which was identified as suitable owing to its favorable stellar density, negligible bright-star contamination, low extinction, and stable visibility \citep{ling2026}. By replicating the planned calibration observations, we ensure that CSST-PSFNet learns from data representative of actual on-orbit conditions.

\begin{figure}
\centering
  \includegraphics[width=0.9\textwidth]{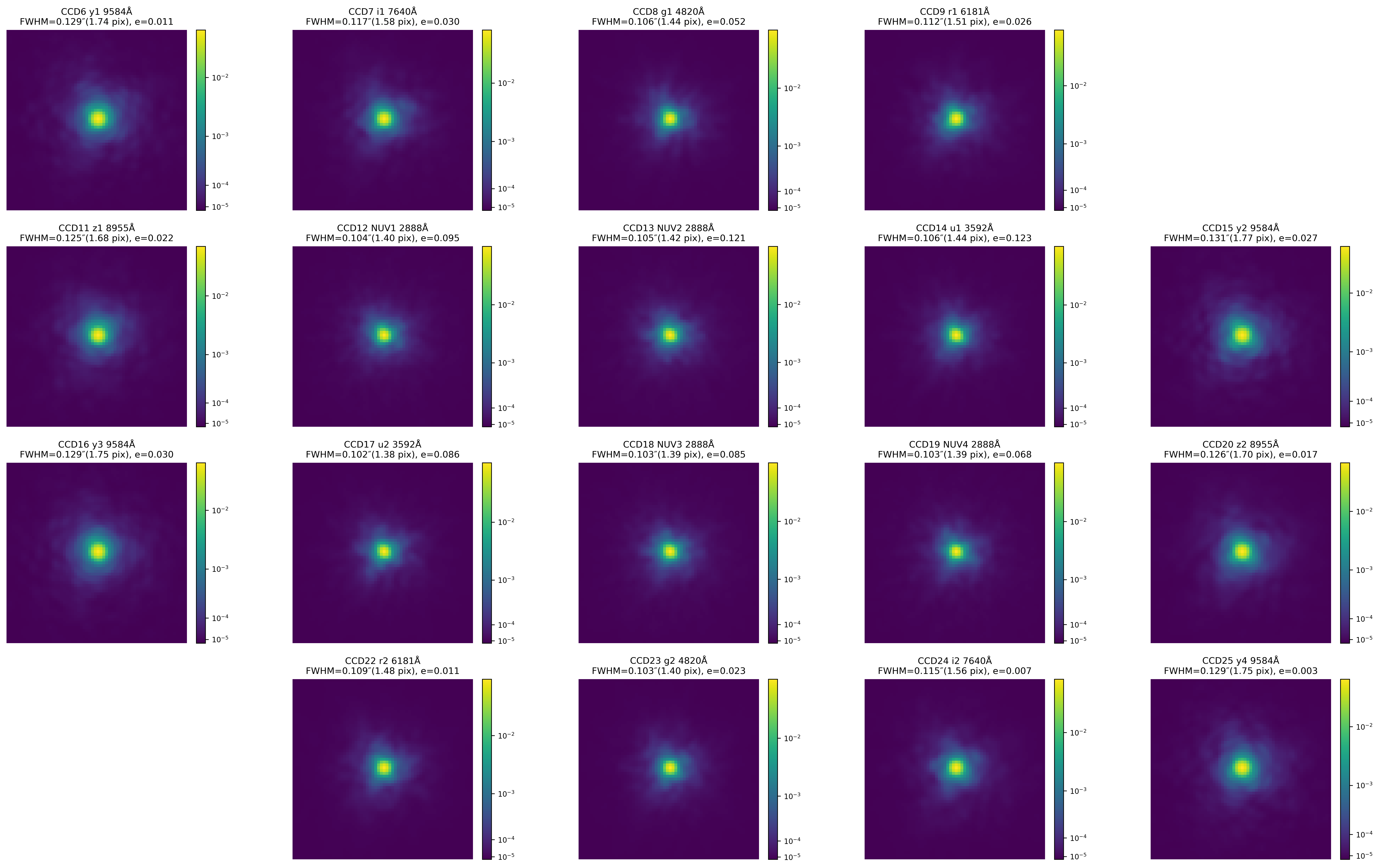}
  \caption{
  Representative PSFs at the detector center from all 18 photometric CCDs in \texttt{DATASET 2}.
  For each CCD, the PSF is extracted from the ground-truth spatial PSF field of \texttt{DATASET 2} at the grid position closest to the CCD center.
  The labels above each subfigure indicate the CCD index, photometric band, effective wavelength, FWHM, and PSF ellipticity $e$.
  The full spatially varying PSF fields in \texttt{DATASET 2} are used as the ground truth for evaluating the corresponding PSF reconstruction results.
  The color scale represents the normalized PSF intensity and is displayed on a logarithmic scale with logarithmically spaced color-bar ticks, highlighting both the PSF core and low-intensity wing structures.
  Note that the displayed PSFs are shown on the $2\times$ upsampled grid (\SI{0.037}{\arcsecond}/pixel) for visualization, but the quoted FWHM values are reported in native CSST pixels (\SI{0.074}{\arcsecond}/pixel) for consistency with instrument specifications; equivalently, the FWHM on the displayed grid is twice the quoted value.
  }
  \label{fig:ground_truth_test_PSF}
\end{figure}

The simulated observations are centered at RA=102.2451$^\circ$ and Dec=-36.0053$^\circ$, providing full coverage across all 18 photometric CCDs.
As shown in Figure~\ref{fig:ground_truth_test_PSF}, the CSST focal plane has a fixed multi-band layout with the central six CCDs corresponding to the $NUV$ and $u$ filters. 
With this pointing, the NGC 2298 core projects primarily between CCD-13 and CCD-18, such that only these two CCDs directly sample the densest cluster region, while the remaining CCDs probe less crowded areas with more uniform stellar distributions. 
For an old, metal-poor globular cluster such as NGC~2298, only a small fraction of stars emit significant flux at short wavelengths, intrinsically limiting detectable sources in the $NUV/u$ bands.

Source detection and stamp extraction are performed using Source Extractor \citep{Bertin_1996}.
Stars are selected based on \texttt{SNR\_WIN}, which represents the signal-to-noise ratio measured within a fixed window centered on the stellar centroid, and the \texttt{FWHM} (full width at half maximum). We apply the criteria $50 \leq \texttt{SNR\_WIN} \leq 300$ and $0 \leq \texttt{FWHM} \leq 4$ pixels and exclude sources flagged as blended or deblended to ensure all samples represent isolated point sources. This yields 17,968 stellar samples in simulated NGC 2298, randomly partitioned into 14,240 training samples (80\%) and 3,728 validation samples (20\%).
Each sample consists of a $32\times32$ stellar cutout, the corresponding $64\times64$ ground truth PSF stamp, and metadata including CCD identifier (iCCD) and focal-plane coordinates $(x,y)$.

To enable controlled evaluation, \texttt{DATASET 2} was generated using the same simulator configuration but an independent source catalog.
Stars were placed on the $30\times30$ PSF grid nodes on each CCD, with flux scaled to achieve $\mathrm{S/N}\approx200$ in the $g$ band and uniform brightness enforced.
This dataset contains approximately 900 samples per CCD and serves as the nominal benchmark for assessing PSF reconstruction accuracy, exhibiting the expected wavelength-dependent morphology (Figure~\ref{fig:ground_truth_test_PSF}).

\begin{figure}
\centering
  \includegraphics[width=0.9\textwidth]{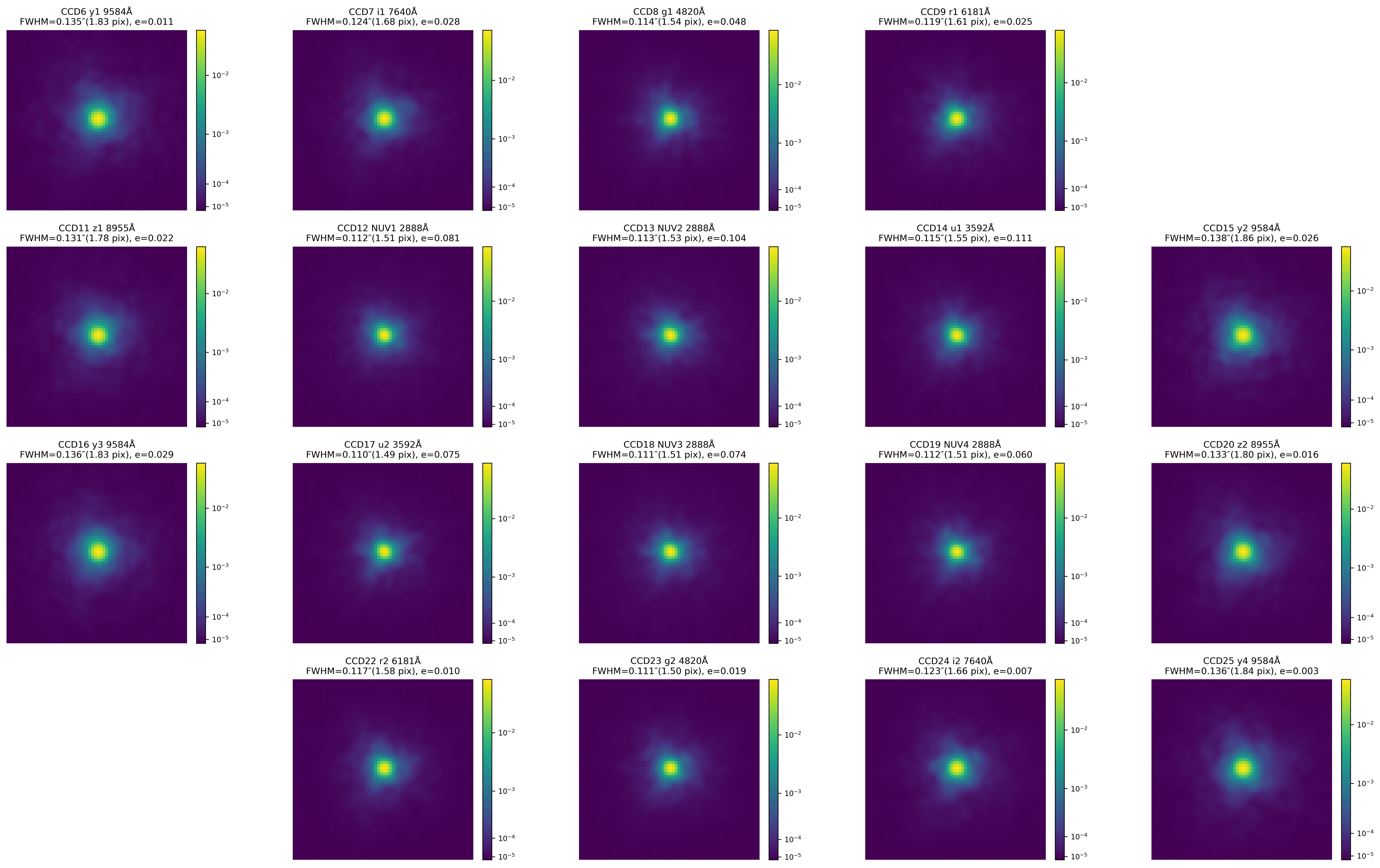}
  \caption{
  Same as Figure~\ref{fig:ground_truth_test_PSF}, but for \texttt{DATASET 3}. Representative PSFs at the detector center are extracted from the ground-truth spatial PSF fields of \texttt{DATASET 3} and used to assess the PSF reconstruction performance under this dataset. The color scale represents the normalized PSF intensity and is shown on a logarithmic scale with logarithmically spaced color-bar ticks, enabling visualization of both the PSF core and faint wing structures.}
  \label{fig:blurred_PSF}
\end{figure}

Finally, \texttt{DATASET 3} was constructed by convolving each PSF in \texttt{DATASET 2} with an isotropic Gaussian kernel, introducing a controlled and simplified PSF broadening scenario. This experiment is intended to represent moderate, approximately isotropic degradation dominated by mild defocus, rather than to capture arbitrary on-orbit PSF perturbations.
The kernel width of $\sigma = 0.5$ pixels is chosen based on OMT simulations of the CSST, which is consistent with typical OMT-predicted defocus amplitudes.
The dataset preserves the same sample count and structure as \texttt{DATASET 2}, while exhibiting broader and more symmetric PSFs consistent with this controlled degradation setting.
Figure~\ref{fig:blurred_PSF} presents the Gaussian-blurred counterparts of the PSFs shown in Figure~\ref{fig:ground_truth_test_PSF}, illustrating the expected increase in FWHM and the corresponding reduction in ellipticity resulting from isotropic convolution.

\section{Model Design and Modeling}
\label{sec: methodology}

\subsection{Model Design}
\begin{figure*}[htbp]
\centering
	\includegraphics[width=\textwidth]{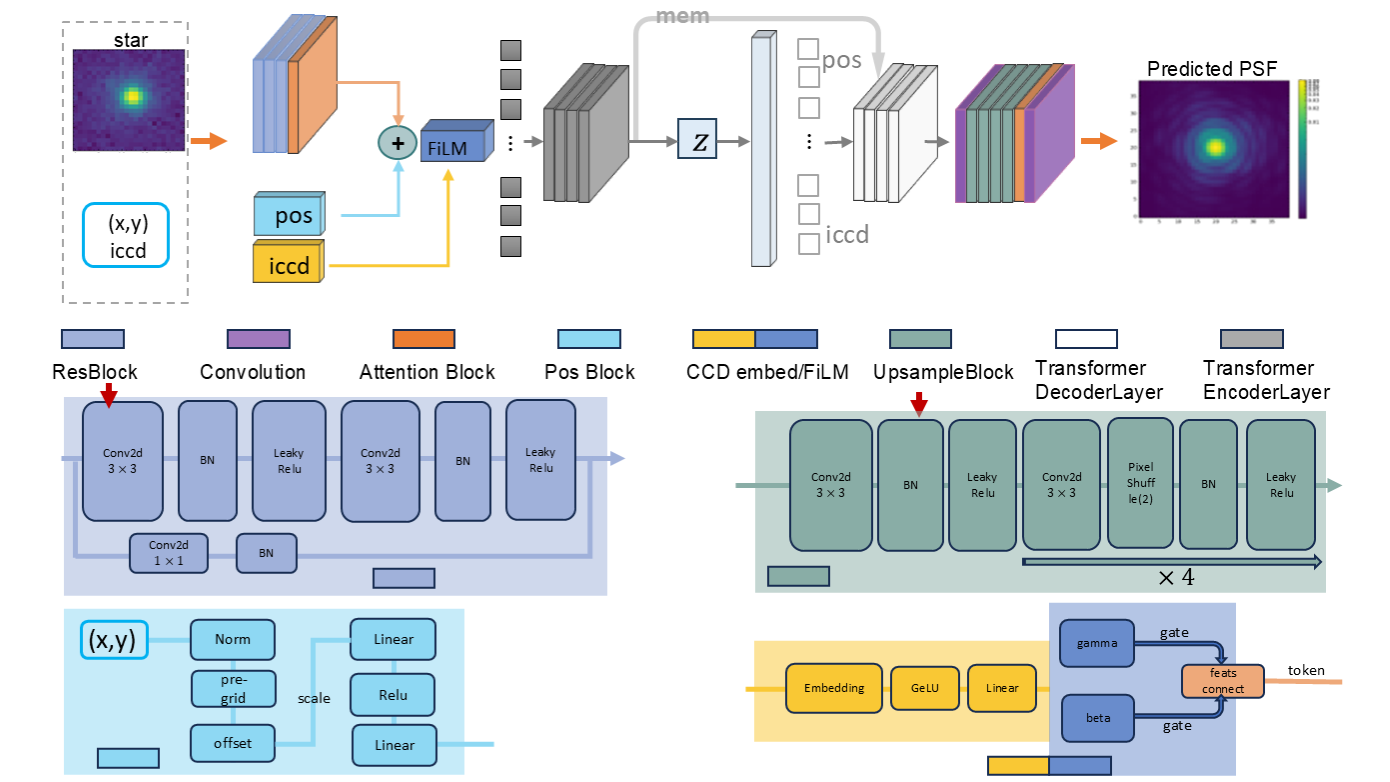}
    \caption{Architecture of \texttt{CSST-PSFNet}, integrating convolutional encoding, CCD/position conditioning, and Transformer-based dual-query decoding for PSF reconstruction.
    Bottom: module details of ResBlock, Pos Block, CCD embed/FiLM, and UpsampleBlock.}
    \label{fig:model_architecture}
\end{figure*}

We propose \texttt{CSST-PSFNet}, a Transformer-augmented conditional variational encoder–decoder specifically designed for instrument-aware PSF super-resolution across the CSST focal-plane array, as illustrated in Figure~\ref{fig:model_architecture}.
The network maps a low-resolution $32\times32$ stellar cutout $\mathbf{x}$, together with the CCD index (iCCD) and normalized focal-plane coordinates $\mathbf{p}$, to a high-fidelity $64\times64$ PSF reconstruction $\mathbf{\hat{y}}$. 

The encoder compresses the stellar input through three residual convolutional modules that capture both the PSF core intensity and diffraction-ring morphology.
A lightweight attention module performs channel–spatial reweighting at the bottleneck scale, enhancing feature coherence and emphasizing physically meaningful structures.
Positional and detector conditions are embedded jointly to provide instrument-aware contextual information.
The encoded features are tokenized and passed through a multi-layer Transformer encoder that models global dependencies across spatial and detector domains.
A variational head projects the encoder output into the parameters of a diagonal Gaussian distribution, from which a latent code $\mathbf{z}$ is sampled using the reparameterization trick.

The decoder performs conditional reconstruction by integrating the latent code $\mathbf{z}$ with positional and detector context. Through cross-attention, the Transformer decoder retrieves PSF features consistent with both spatial position and CCD-specific properties. 
The decoded representation is progressively upsampled through a lightweight reconstruction stack to produce the $64\times64$ high-resolution PSF.
Upsampling is implemented using sub-pixel convolution (PixelShuffle), which mitigates aliasing and preserves structural continuity.
A shallow convolutional head maps the final feature maps into a single-channel PSF image. Non-negative intensities are enforced by a sigmoid activation at the output layer. The reconstructed PSF is then normalized by its total flux on a per-sample basis to ensure unit flux.

Accurate PSF reconstruction under the severe undersampling of CSST requires the network to preserve both local morphological details and global photometric consistency.
The residual and attention modules provide strong local representation capability, while the Transformer encoder captures long-range dependencies and detector-level correlations.
To represent smooth spatial variations and inter-detector differences, the positional encoder transforms focal-plane coordinates $(x,y)$ into learnable geometric features, and the CCD embedding encodes detector-specific statistics.

The model is optimized with a composite loss that includes pixel reconstruction and variational regularization.
Let $\hat{\mathbf{P}}$, $\mathbf{P}\in\mathbb{R}^{H\times W}$ denote the reconstructed and ground-truth PSFs, respectively.
The total loss is
\begin{equation}
\label{eq:total_loss}
\mathcal{L}_{\mathrm{total}}
=\lambda_{\mathrm{rec}}\mathcal{L}_{\mathrm{rec}}
+\lambda_{\mathrm{kld}}\mathcal{L}_{\mathrm{kld}} .
\end{equation}

Reconstruction loss measures the fidelity at the pixel-level between $\hat{\mathbf{P}}$ and $\mathbf{P}$,
\begin{equation}
\label{eq:recon_loss}
\mathcal{L}_{\mathrm{rec}}
=\|\hat{\mathbf{P}}-\mathbf{P}\|_F^2
=\sum_{u=1}^{H}\sum_{v=1}^{W}
(\hat{P}_{uv}-P_{uv})^2 .
\end{equation}

The variational regularization term constrains the latent posterior
$q(\mathbf{z}|\mathbf{x})=\mathcal{N}(\boldsymbol{\mu},\mathrm{diag}(\boldsymbol{\sigma}^2))$
toward the unit Gaussian prior
$p(\mathbf{z})=\mathcal{N}(\mathbf{0},\mathbf{I})$
via the Kullback–Leibler divergence,
\begin{equation}
\mathcal{L}_{\mathrm{kld}}
=\tfrac{1}{2}\mathbb{E}\left[
\|\boldsymbol{\mu}\|2^2
+\|\boldsymbol{\sigma}\|2^2
-\log \boldsymbol{\sigma}^2 - 1
\right].
\end{equation}
Its weight $\lambda_{\mathrm{kld}}$ is linearly annealed during the warm-up phase,
\begin{equation}
\lambda_{\mathrm{kld}}(t)
=\lambda{\min}
+(\lambda_{\max}-\lambda_{\min})
\min\left(\frac{t}{T_{\mathrm{ramp}}},1\right),
\end{equation}
where $t$ is the epoch index and
($\lambda_{\min},\lambda_{\max},T_{\mathrm{ramp}}$) are the annealing parameters.

\subsection{Modeling}
\label{sec: modeling} 
\texttt{CSST-PSFNet} is trained using the AdamW optimizer with an initial learning rate of $1\times10^{-4}$ and a weight decay of $1\times10^{-5}$. A five-epoch linear warm-up is followed by a cosine decay schedule that gradually reduces the learning rate to 1\% of its initial value. Training proceeds for up to 200 epochs with early stopping (patience of 20), and convergence is typically reached at epoch 40. We adopt a batch size of 16, employ mixed-precision training, and apply gradient clipping at~0.5 to ensure stable optimization. We verified that moderate variations of learning rate and latent dimension do not significantly affect the reconstruction metrics. The training objective combines a reconstruction term and a KLD regularization term, with weights $\lambda_{\mathrm{rec}}=1.0$ and $\lambda_{\mathrm{kld}}=1.0$, respectively.
All experiments are conducted on a Linux server equipped with an NVIDIA A100 GPU (40 GB). The implementation is based on PyTorch and is executed in a CUDA 12.2 runtime environment. 

\texttt{DATASET 1} is split into 14,240 training samples (about 890 batches) and 3,728 samples (about 223 batches) for validation. Both evaluation datasets, \texttt{DATASET 2} and \texttt{DATASET 3}, comprise approximately 860-900 stellar samples per CCD over all 18 imaging CCDs. Whereas \texttt{DATASET 2} serves as the nominal reference set, \texttt{DATASET 3} augments it by applying an isotropic Gaussian kernel to each PSF, producing a systematically broadened counterpart for testing robustness to realistic on-orbit perturbations.

\subsection{Evaluation Metrics and Validation}
\label{sec:metrics}
To quantitatively evaluate the fidelity of PSF reconstruction, we adopt two complementary categories of metrics: 
(1) pixel-level image fidelity indices, which measure the similarity between reconstructed and reference PSF images at the pixel scale, and 
(2) shear-related shape metrics, which probe the recovery of PSF size and ellipticity---quantities that are of central importance for weak-lensing cosmology.

\subsubsection{Pixel-level metrics}
\label{sec:pixel_metrics}
Pixel-level metrics provide a direct evaluation of the similarity between the reconstructed PSF image $I_{\rm rec}$ 
and the ground-truth PSF $I_{\rm true}$. We consider three complementary measures:

1. \textbf{Root Mean Square Error (RMSE);}  
 RMSE quantifies the average pixel-wise deviation between two images and is defined as
    \begin{equation}
    {\rm RMSE} = \sqrt{\frac{1}{N} \sum_{i=1}^N \left(I_{{\rm rec},i} - I_{{\rm true},i}\right)^2},
    \end{equation}
    where $N$ is the number of pixels.  
    It is reported in the same units as the image intensity and reflects the typical reconstruction error per pixel. 
    In PSF modeling, a smaller RMSE indicates closer agreement with the reference, which serves as a baseline measure of pixel-level reconstruction fidelity.

2. \textbf{Peak Signal-to-Noise Ratio (PSNR).}  
    PSNR expresses the reconstruction quality on a logarithmic scale:
    \begin{equation}
    {\rm PSNR} = 20 \log_{10}\left(\frac{I_{\rm max}}{\sqrt{{\rm MSE}}}\right),
    \quad {\rm MSE} = \frac{1}{N}\sum_{i=1}^N \left(I_{{\rm rec},i} - I_{{\rm true},i}\right)^2,
    \end{equation}
    where $I_{\rm max}$ is the maximum intensity in the ground truth PSF.  
    Higher PSNR values indicate superior fidelity to the reference PSF.  
    
3. \textbf{Structural Similarity Index (SSIM).}  
    SSIM measures image similarity in terms of luminance, contrast, and structural information \citep{Wang_2004, Avanaki_2009}:
    \begin{equation}
    {\rm SSIM}(x,y) = \frac{(2\mu_x\mu_y+C_1)(2\sigma_{xy}+C_2)}
    {(\mu_x^2+\mu_y^2+C_1)(\sigma_x^2+\sigma_y^2+C_2)},
    \end{equation}
    where $\mu_x,\mu_y$ are the local means, $\sigma_x^2,\sigma_y^2$ are the variances and $\sigma_{xy}$ is the covariance. The constants $C_1$ and $C_2$ are included to ensure numerical stability in low-intensity regions of PSF images, where the local mean and variance can approach zero. SSIM ranges from 0 to 1, with higher values denoting greater structural similarity.

\subsubsection{Shear-related metrics}
\label{sec:shape_metrics}
Besides pixel-level fidelity, weak-lensing science places stringent requirements on the recovery of PSF size and shape, 
since residual errors in these quantities propagate directly into multiplicative and additive shear biases 
\citep{Paulin_Henriksson_2008, Paulin_Henriksson_2009}. 
Therefore, we employ the adaptive-moment formalism \citep{Hirata_2003} to assess the precision of reconstructed PSFs. 
In this approach, the surface brightness distribution $I(x,y)$ is characterized by its adaptive second moments,
measured with an elliptical Gaussian weight function, iteratively matched to the object profile. 

The adaptive quadrupole moments are defined as
\begin{equation}
Q_{ij} = \frac{\int (x_i - \bar{x}_i)(x_j - \bar{x}_j) I(x,y) W(x,y)\, dxdy}{\int I(x,y)W(x,y)\,dxdy}, \quad i,j \in \{x,y\},
\end{equation}
where $W(x,y)$ is the elliptical Gaussian weight matched and $(\bar{x}_i,\bar{x}_j)$ is the centroid. 
From these moments, the effective PSF size is given by
\begin{equation}
R^2 = Q_{xx} + Q_{yy},
\end{equation}
and the ellipticity components are expressed as
\begin{equation}
e_1 = \frac{Q_{xx}-Q_{yy}}{Q_{xx}+Q_{yy}}, 
\qquad
e_2 = \frac{2Q_{xy}}{Q_{xx}+Q_{yy}},
\qquad
e = \sqrt{e_1^2+e_2^2}.
\end{equation}

We define the size and ellipticity deviation as follows:
\begin{equation}
\delta R^2/R^2 = \frac{R^2_{\rm rec}-R^2_{\rm true}}{R^2_{\rm true}}, 
\qquad
\delta e = e_{{\rm rec}} - e_{{\rm true}}, \; 
\end{equation}
where the subscripts 'rec' and 'true' refer to measurements from the reconstructed PSFs (either \texttt{PSFEx} or \texttt{CSST-PSFNet}) and the corresponding ground-truth PSFs, respectively.
\cite{Heymans_2005} modeled the time and position dependent variation of the PSF in the absence of a known true PSF by fitting the stellar ellipticity field across the HST/ACS images, and evaluated the model accuracy using the root-mean-square (RMS) of the stellar–model ellipticity residuals. 
In this work, since the simulated datasets provide the theoretical PSF as ground truth, we can directly quantify the reconstruction accuracy by computing the standard deviations of the residuals in both shape and size parameters:
\[
\sigma({\delta e}) = \mathrm{std}(\delta e), \qquad
\sigma({\delta R^2 / R^2}) = \mathrm{std}\left(\frac{\delta R^2}{R^2_{\mathrm{true}}}\right).
\]
These standard deviation metrics are numerically equivalent to the stellar residual RMS used in weak-lensing PSF studies (e.g., \citealt{Heymans_2005}) and provide a proxy to the additive and multiplicative shear calibration biases potentially introduced by PSF reconstruction errors.

\section{Performance Analysis}
\label{sec: experiments}
To contextualize the performance of \texttt{CSST-PSFNet}, we adopt \texttt{PSFEx} as a reference baseline. \texttt{PSFEx} is a survey-grade PSF modeling method widely used in weak-lensing studies and validated across major surveys such as HSC \citep{Mandelbaum_2017}, DES \citep{Zuntz_2018}, and CFIS \citep{Guinot_2022}.
Working jointly with \texttt{Source Extractor} \citep{Bertin_1996}, it constructs spatially varying PSF models from stellar cutouts using a non-parametric formulation \citep{Bertin_2011}.
Each PSF is expressed as a linear combination of eigen-PSFs derived from singular value decomposition, while the spatial variation of the associated coefficients is captured through smooth low-order polynomials across the focal plane. 

In this study, \texttt{PSFEx} is configured with a third-order polynomial spatial model and produces PSFs on a $2\times$ oversampled grid, which matches the target resolution of \texttt{CSST-PSFNet}. A key methodological distinction is that \texttt{CSST-PSFNet} is trained once on \texttt{DATASET 1} and applied to all test sets with fixed parameters, whereas PSFEx is re-fitted independently on each dataset, reflecting different operational paradigms.
All data preprocessing steps are kept consistent with those used for \texttt{CSST-PSFNet}, ensuring that the comparison isolates modeling performance rather than differences in data handling.

\subsection{Quantitative Results on \texttt{DATASET 2}}
\begin{figure}[h]
  \centering
    \includegraphics[width=0.8\textwidth]{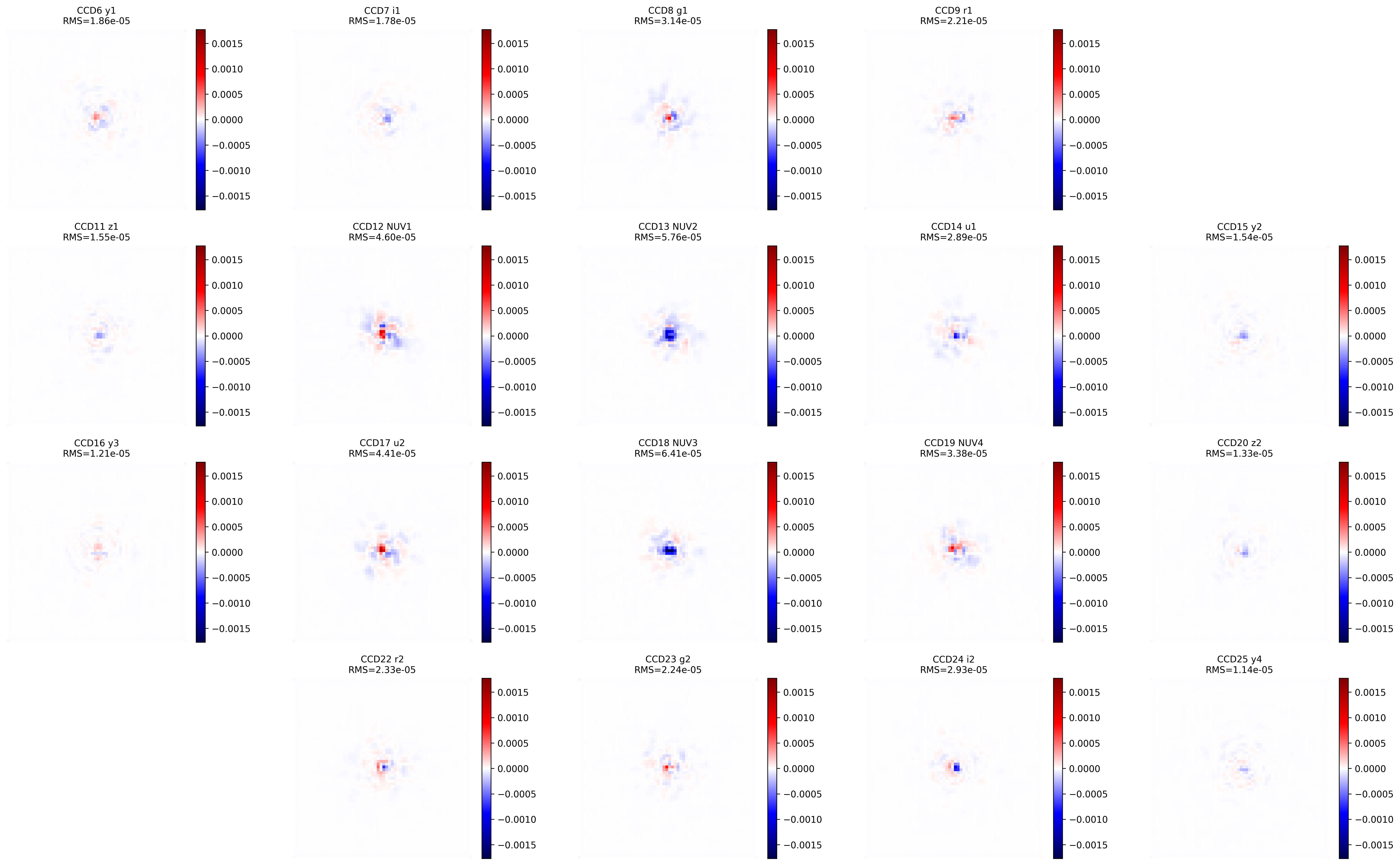}
    \caption{
    Residual maps of the central PSFs reconstructed by \texttt{CSST-PSFNet} for all 18 CCDs on \texttt{DATASET 2}. 
    Compared to the \texttt{PSFEx} results (Figure~\ref{fig:psfex_residual}), the model residuals are typically an order of magnitude smaller in RMS ($\sim10^{-5}$ versus $\sim10^{-4}$), and large-scale ring-like artifacts are absent. 
    Only fine high-frequency structures of low amplitude remain, demonstrating the model’s improved ability to capture both PSF cores and wings across the full focal plane.
    }
    \label{fig:model_residual}
\end{figure}

\begin{figure}[h]
  \centering
    \includegraphics[width=0.8\textwidth]{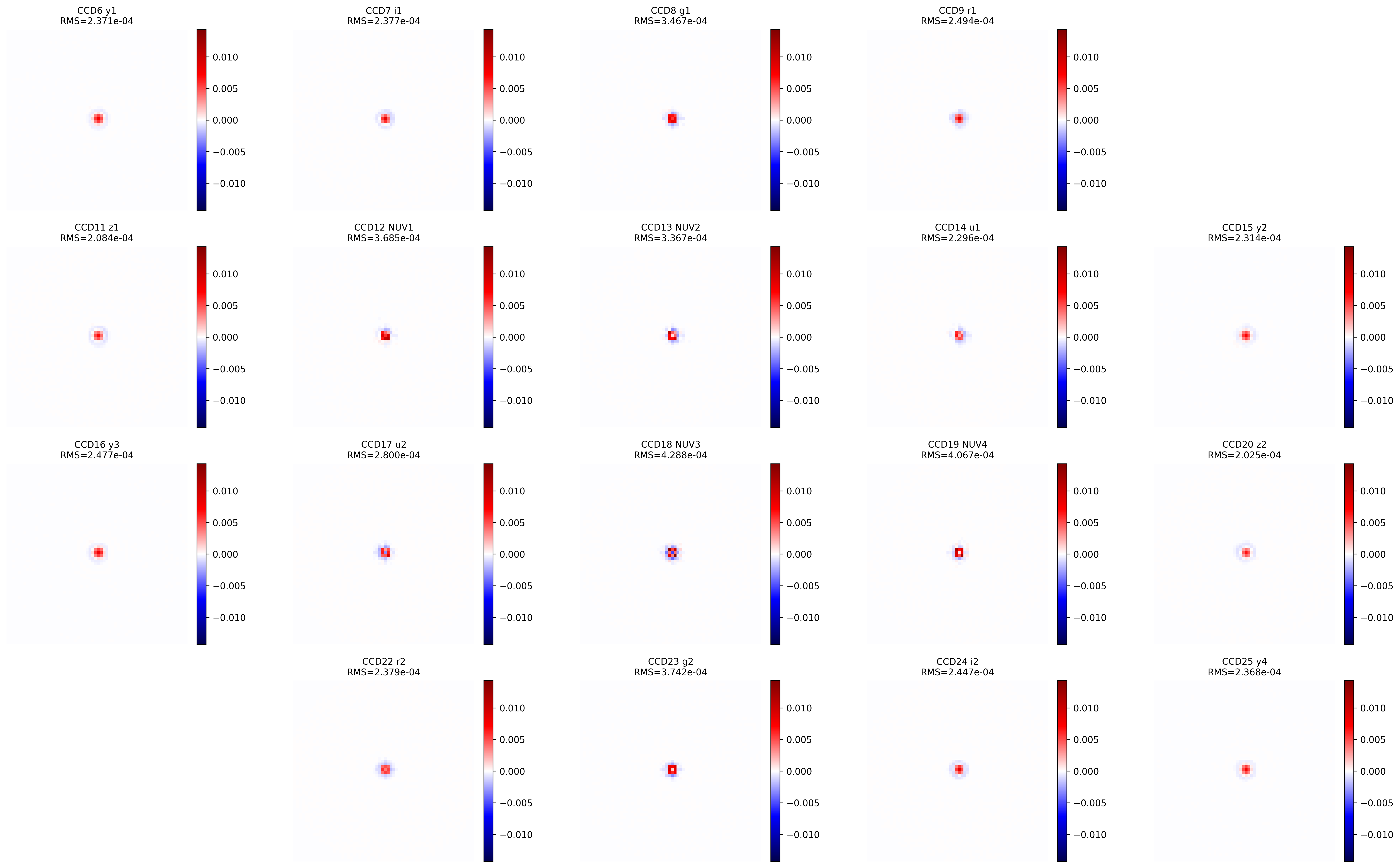}
    \caption{
    Residual maps of the central PSFs reconstructed by \texttt{PSFEx} for all 18 CCDs on \texttt{DATASET 2}, relative to the ground truth PSFs shown in Figure~\ref{fig:ground_truth_test_PSF}. The reconstructions reproduce the overall morphology but leave prominent features, most notably bright central cores (red) surrounded by negative rings (blue).  These features indicate difficulties in simultaneously matching both the PSF core and extended wings.}
    \label{fig:psfex_residual}
\end{figure}

We begin our comparison with residual maps across the full focal plane. For the same positions shown in Figure~\ref{fig:ground_truth_test_PSF},  Figures~\ref{fig:model_residual} and \ref{fig:psfex_residual} display the residuals of the \texttt{CSST-PSFNet} and  \texttt{PSFEx} reconstructions. Both methods reproduce the gross morphology, but their residual patterns differ substantially. \texttt{PSFEx} solutions typically show a bright positive core surrounded by a negative ring (e.g., CCD 6–7, 9, 11, 15–16, 20, 24–25), and in the $NUV$ and $u$ bands exhibit concentric positive–negative rings, reflecting the difficulty in simultaneously fitting the PSF core and extended wings. In contrast,  \texttt{CSST-PSFNet} residuals are typically an order of magnitude smaller (RMS $\sim10^{-5}$ v.s. $\sim10^{-4}$ for \texttt{PSFEx}) and do not display such large-scale ring-like artifacts. Instead, they contain only fine, low-amplitude high-frequency structures, indicating that while large-scale residuals are strongly suppressed, small-scale deviations remain. These results highlight the ability of \texttt{CSST-PSFNet} to simultaneously capture the PSF core and wings, a requirement of weak-lensing science. 

\begin{figure}[h]
  \centering
  \begin{subfigure}{0.45\textwidth}
    \includegraphics[width=\textwidth]{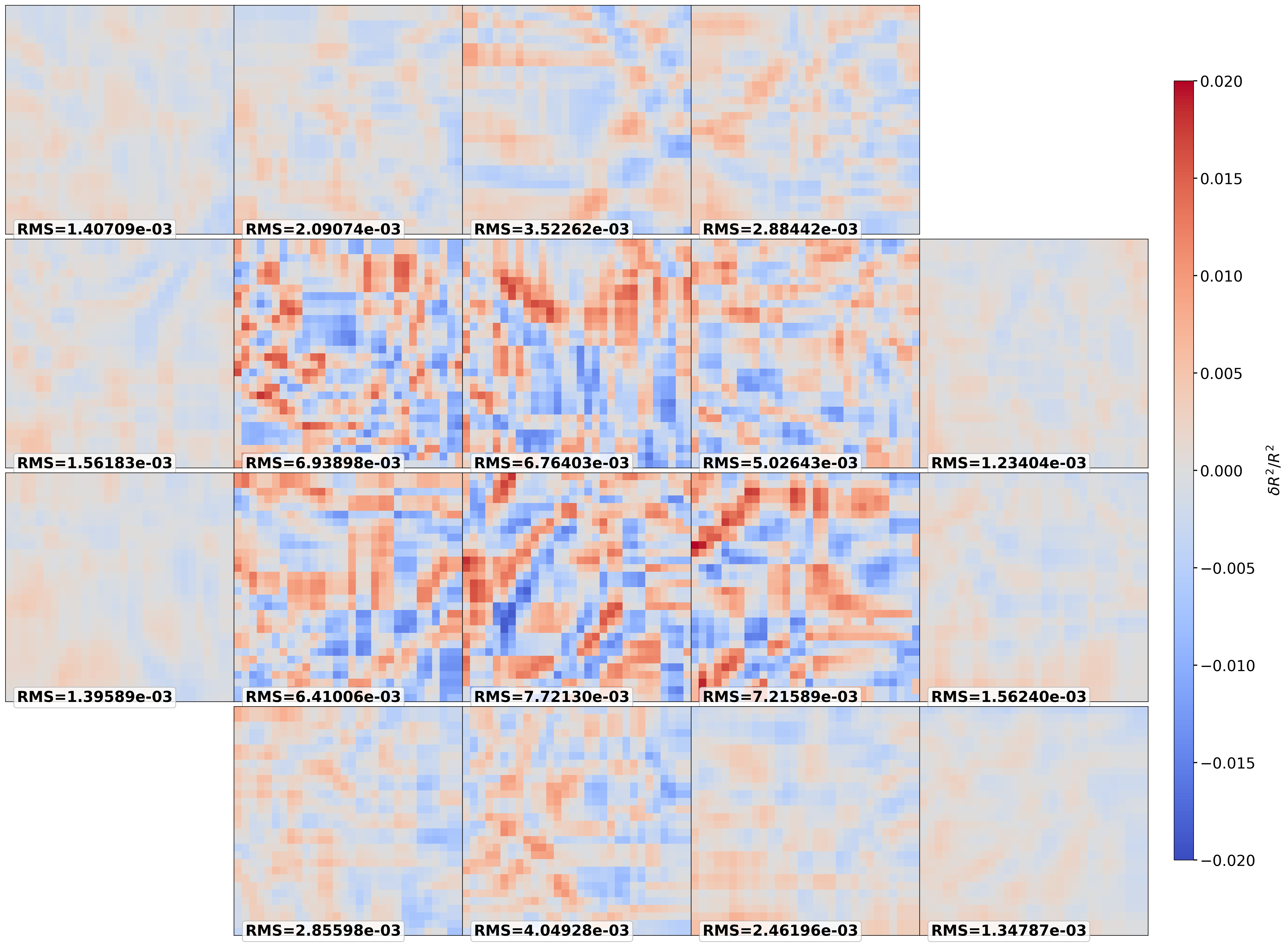}
    \caption{$\delta R^2/R^2$ distribution of \texttt{DATASET 2}}
    \label{Model.Residual.R.SNR200}
  \end{subfigure}
  \hfill
  \begin{subfigure}{0.45\textwidth}
    \includegraphics[width=\textwidth]{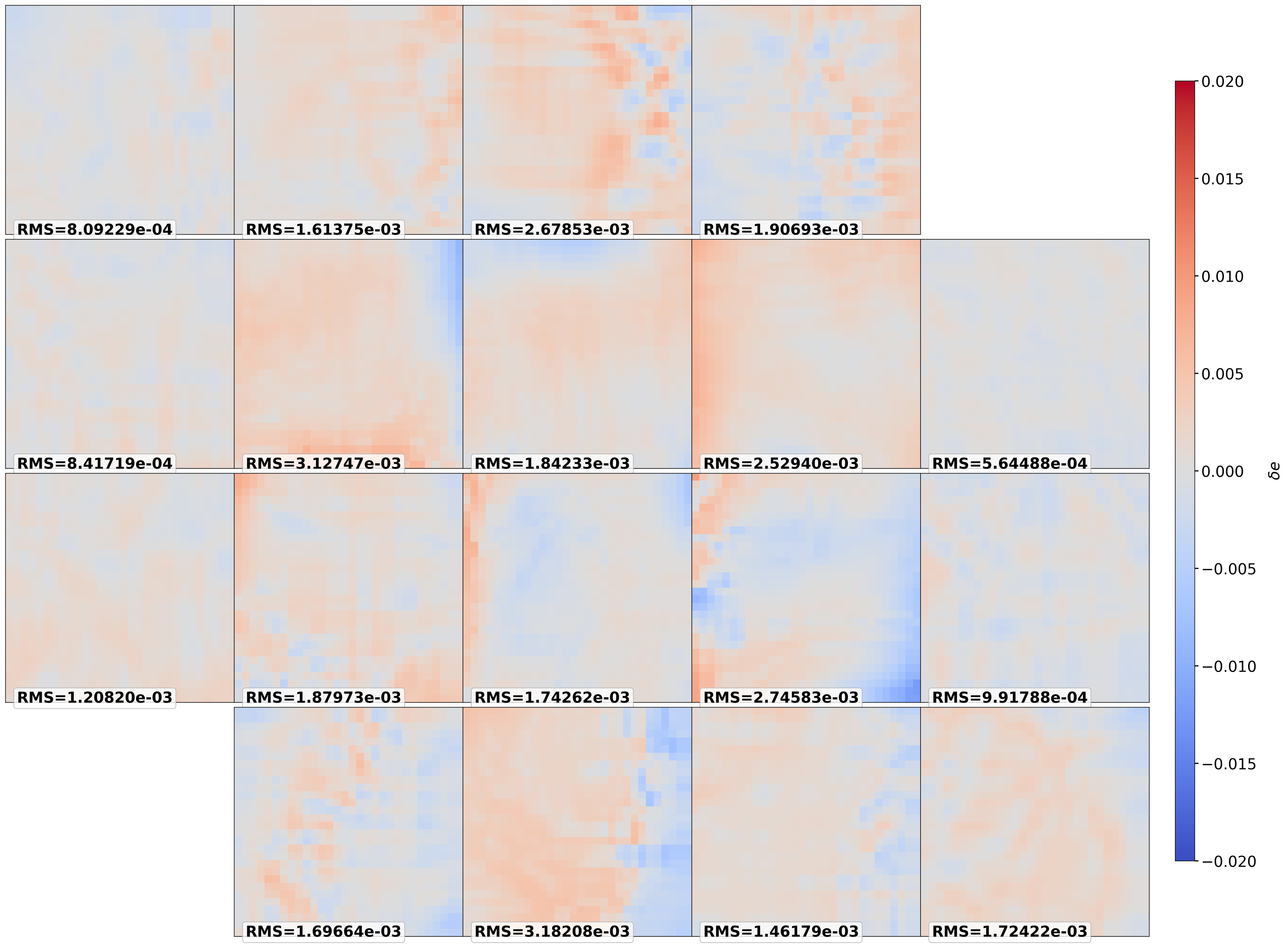}
    \caption{$\delta e$ distribution of \texttt{DATASET 2}}
    \label{Model.Residual.e.SNR200}
  \end{subfigure}
  \caption{Residual distributions of PSF size ($\delta R^2/R^2$, panel a) and ellipticity ($\delta e$, panel b) across all 18 CCDs for  \texttt{CSST-PSFNet}, evaluated on \texttt{DATASET 2}. 
    The residuals remain compact and spatially uniform, with RMS values at the $10^{-3}$ level and deviations generally confined within $\pm0.5\%$ for PSF size and $\sim10^{-3}$ for ellipticity. 
    No large-scale coherent artifacts are visible, indicating that  \texttt{CSST-PSFNet} accurately captures both PSF size and anisotropy across the focal plane.}
  \label{fig:Model.Residual_R_e.SNR200}
\end{figure}

\begin{figure}[h]
  \centering
  \begin{subfigure}{0.48\textwidth}
    \includegraphics[width=\textwidth]{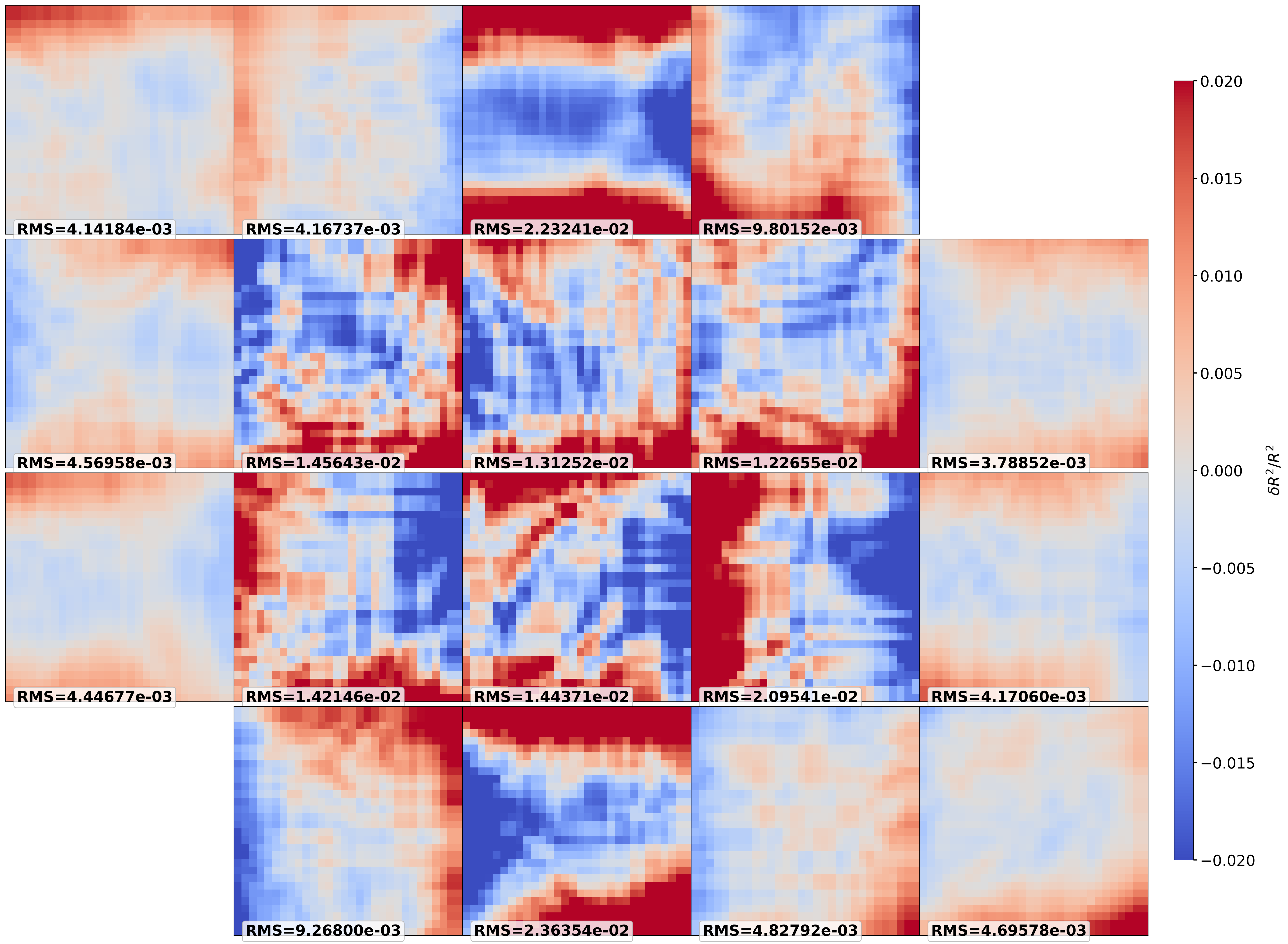}
    \caption{$\delta R^2/R^2$ distribution of \texttt{DATASET 2}}
    \label{PSFEx.Residual.R.SNR200}
  \end{subfigure}
  \hfill
  \begin{subfigure}{0.48\textwidth}
    \includegraphics[width=\textwidth]{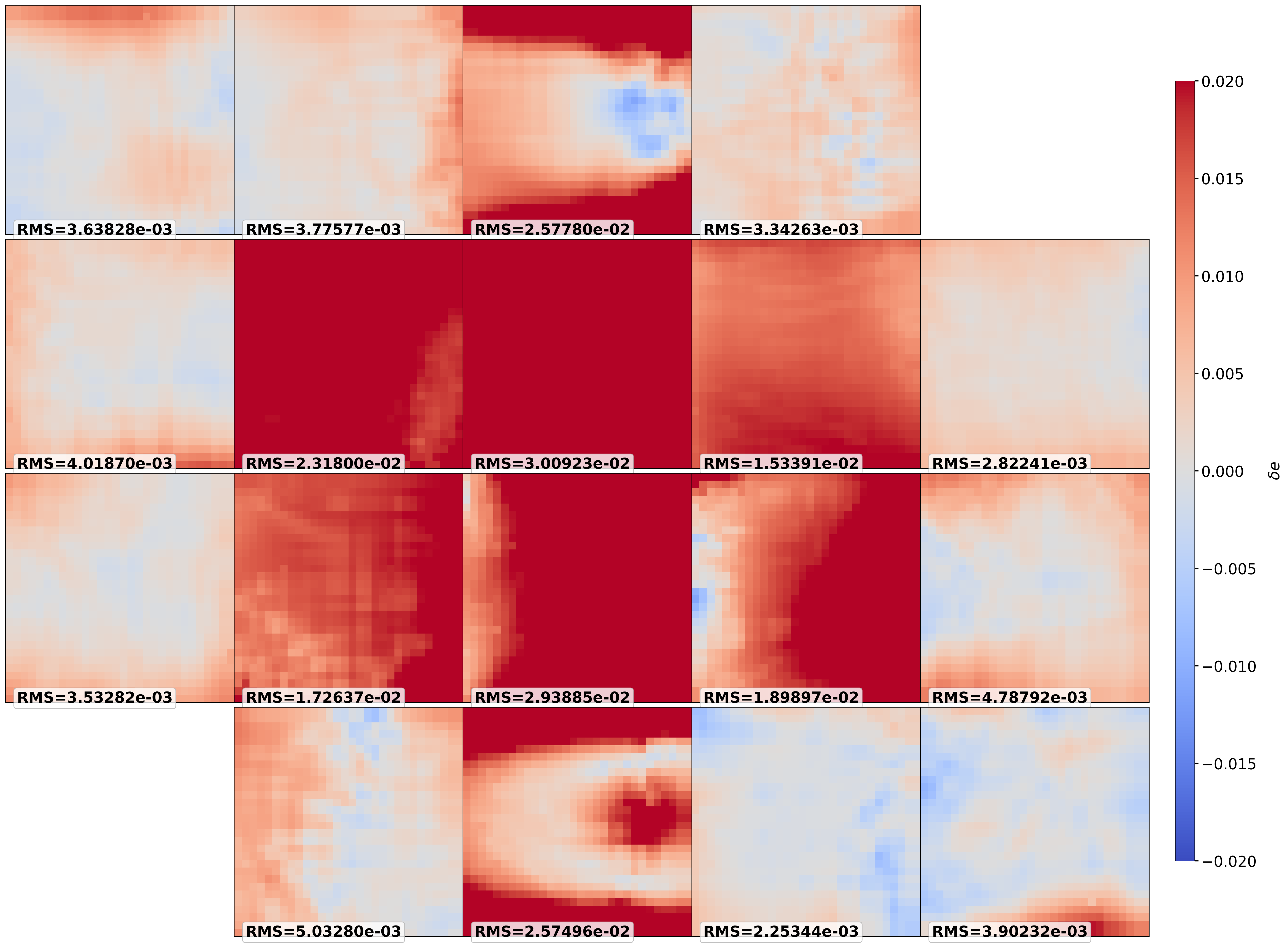}
    \caption{$\delta e$ distribution of \texttt{DATASET 2}}
    \label{PSFEx.Residual.e.SNR200}
  \end{subfigure}
  \caption{
    Residual distributions of PSF size ($\delta R^2/R^2$, panel a) and ellipticity ($\delta e$, panel b) across all 18 CCDs for \texttt{PSFEx}, evaluated on \texttt{DATASET 2}. 
    Compared to  \texttt{CSST-PSFNet}, \texttt{PSFEx} shows substantially larger residuals: the PSF size residuals exhibit stripe-like and patchy patterns with amplitudes of several percent (RMS up to $2\times10^{-2}$), while the ellipticity residuals often exceed the $10^{-2}$ level and display saturated regions close to the $\pm0.02$ limits, particularly in the undersampled $NUV$, $u$, and $g$ bands. 
    These results highlight the limitations of \texttt{PSFEx} in capturing spatial variations of PSF morphology.
    }
  \label{fig:PSFEx.Residual_R_e.SNR200}
\end{figure}

In addition to visual inspection, we analyzed the spatial distributions of the FWHM and ellipticity residuals throughout the entire focal plane. Figures~\ref{fig:Model.Residual_R_e.SNR200} and \ref{fig:PSFEx.Residual_R_e.SNR200} show the maps of  $\delta R^2/R^2$ and $\delta e$ for  \texttt{CSST-PSFNet} and for \texttt{PSFEx}.  
For FWHM (Figures~\ref{Model.Residual.R.SNR200}, \ref{PSFEx.Residual.R.SNR200}),  \texttt{CSST-PSFNet} maintains deviations within $\pm0.5\%$, with RMS values at the 0.001 level, and shows no coherent large-scale patterns. In contrast, \texttt{PSFEx} reconstructions exhibit strong alternating positive–negative stripe and patch-like structures, with amplitudes reaching several percent. The effect is particularly severe in the $NUV$ and $u$ bands, where undersampling is strongest in the central six CCDs, and RMS values can reach 0.02.
For ellipticity residuals (Figures~\ref{Model.Residual.e.SNR200}, \ref{PSFEx.Residual.e.SNR200}), \texttt{CSST-PSFNet} consistently achieves low RMS levels of $5.6\times10^{-4}$ -  $3.1\times10^{-3}$ across all CCDs, with spatially uniform distributions and no large-scale artifacts. By contrast, \texttt{PSFEx} frequently exceeds the 0.01 level, with several CCDs in the undersampled $NUV$, $u$, and parts of the $g$ band showing saturated residual patches close to the colorbar limits $\pm0.02$, indicating substantial inaccuracies in modeling the anisotropy of the PSF.
In general,  \texttt{CSST-PSFNet} reduces residuals in both $R$ and $e$ by at least an order of magnitude compared to \texttt{PSFEx}, achieving more accurate and spatially uniform PSF reconstructions. This improvement is critical for weak-lensing applications, where even percent-level biases in PSF size or ellipticity directly propagate into shear measurement residuals. 

Table~\ref{tab:pixel_metrics} summarizes the pixel-level reconstruction quality.
Compared with PSFEx, CSST-PSFNet reduces the RMSE from $\sim3\times10^{-3}$ to $\sim2\times10^{-4}$ (a $\sim$15$\times$ improvement, i.e., just over one order of magnitude),
while the PSNR increases from 51 to 71.
The SSIM values correspondingly move closer to unity, indicating improved structural fidelity.
In particular, these gains are observed throughout the spectral range, from the severely undersampled $NUV$ and $u$
bands to the longer-wavelength $y$ band, demonstrating that the proposed method retains superior reconstruction fidelity, irrespective of the level of undersampling. This consistency across bands emphasizes the robustness of \texttt{CSST-PSFNet}, particularly in challenging, undersampled conditions. It reinforces its capability to produce high-quality reconstructions across the full CSST spectral range. In addition, PSFEx requires approximately 4738 seconds to construct the PSF model for the dataset, whereas CSST-PSFNet completes inference in 296 seconds, corresponding to an approximately 16× reduction in runtime.

\begin{table}
\centering
\caption{Per-band averages of pixel-level reconstruction metrics for \texttt{PSFEx} and \texttt{CSST-PSFNet} evaluated on \texttt{DATASET 2}. 
Lower RMSE and higher PSNR/SSIM indicate better performance.}
\label{tab:pixel_metrics}
\begin{tabular}{c|ccc|ccc}
\hline\hline
 & \multicolumn{3}{c|}{\texttt{PSFEx}} & \multicolumn{3}{c}{Model} \\
Bandpass & PSNR & SSIM & RMSE & PSNR & SSIM & RMSE \\
\hline
NUV & 47.9108 & 0.9979 & 0.0041 & 67.9715 & 0.9998 & 0.0004 \\
u  & 50.5035 & 0.9967 & 0.0030 & 69.6252 & 0.9999 & 0.0003 \\
g & 50.8477 & 0.9962 & 0.0029 & 72.4350 & 0.9999 & 0.0002 \\
r & 53.2417 & 0.9967 & 0.0021 & 72.0391 & 0.9999 & 0.0002 \\
i  & 52.8541 & 0.9986 & 0.0022 & 70.9628 & 0.9999 & 0.0002 \\
z & 53.2098 & 0.9986 & 0.0021 & 73.9183 & 0.9999 & 0.0002 \\
y  & 53.5175 & 0.9990 & 0.0021 & 73.6150 & 0.9999 & 0.0002 \\
\hline
Mean & 51.5014 & 0.9978 & 0.0027 & 71.3503 & 0.9999 & 0.0002 \\
% Std  & 0.001 & 0.5  & 0.007 & 0.001 & 0.6  & 0.005 \\
\hline
\end{tabular}
\end{table}

\subsection{Quantitative Results on \texttt{DATASET 3}}
To evaluate the robustness of \texttt{CSST-PSFNet} under realistic on-orbit perturbations, we compare both methods on \texttt{DATASET 3}. Because the applied Gaussian blur suppresses high-frequency structures, the residual maps of PSF reconstruction become less informative, and therefore, our analysis for \texttt{DATASET 3} focuses on the physically meaningful PSF size and ellipticity residuals.

\begin{figure}[h]
  \centering
  \begin{subfigure}{0.45\textwidth}
    \includegraphics[width=\textwidth]{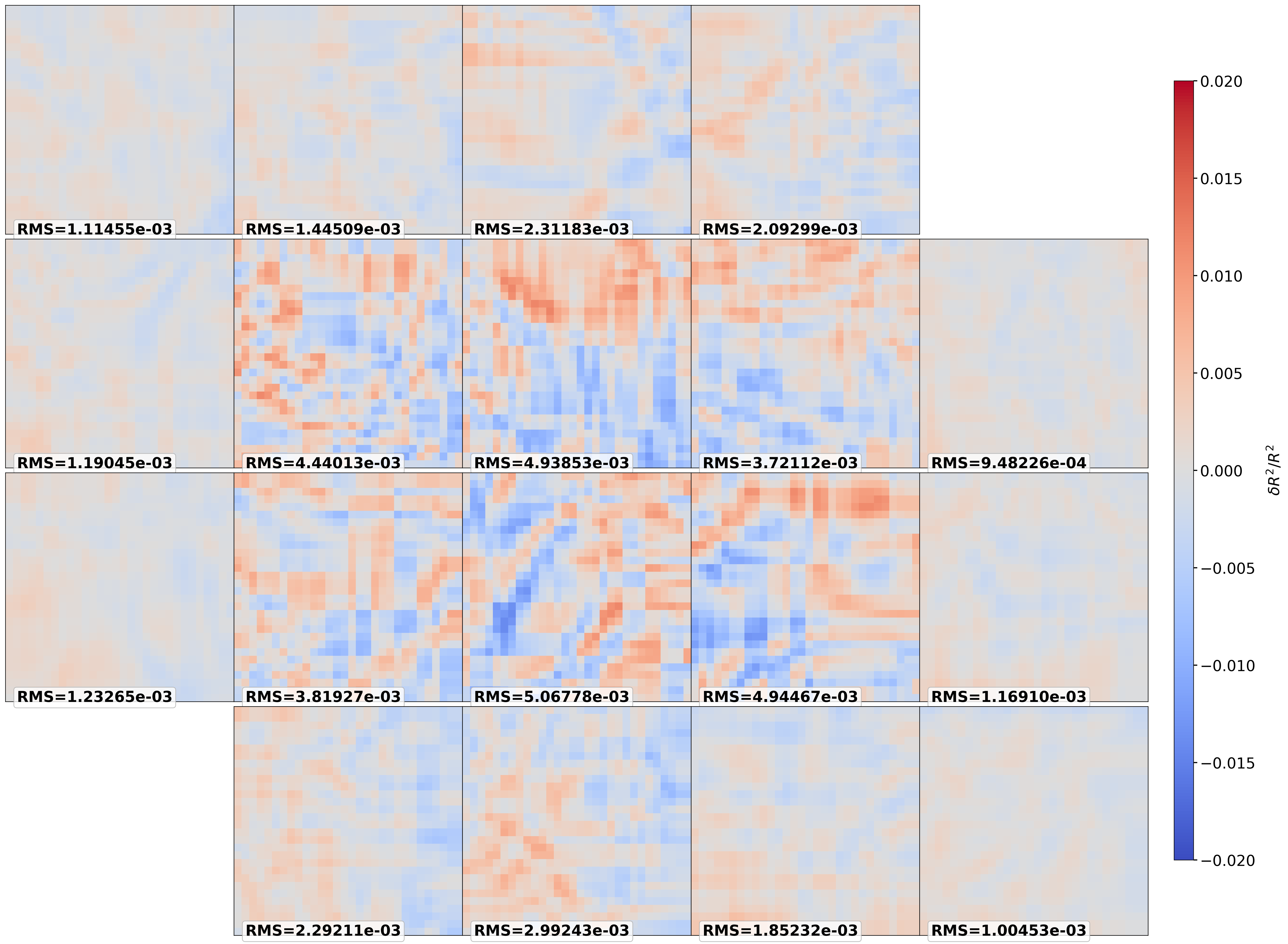}
    \caption{$\delta R^2/R^2$ distribution of \texttt{DATASET 3}}
    \label{Model.Residual.R.blur}
  \end{subfigure}
  \hfill
  \begin{subfigure}{0.45\textwidth}
    \includegraphics[width=\textwidth]{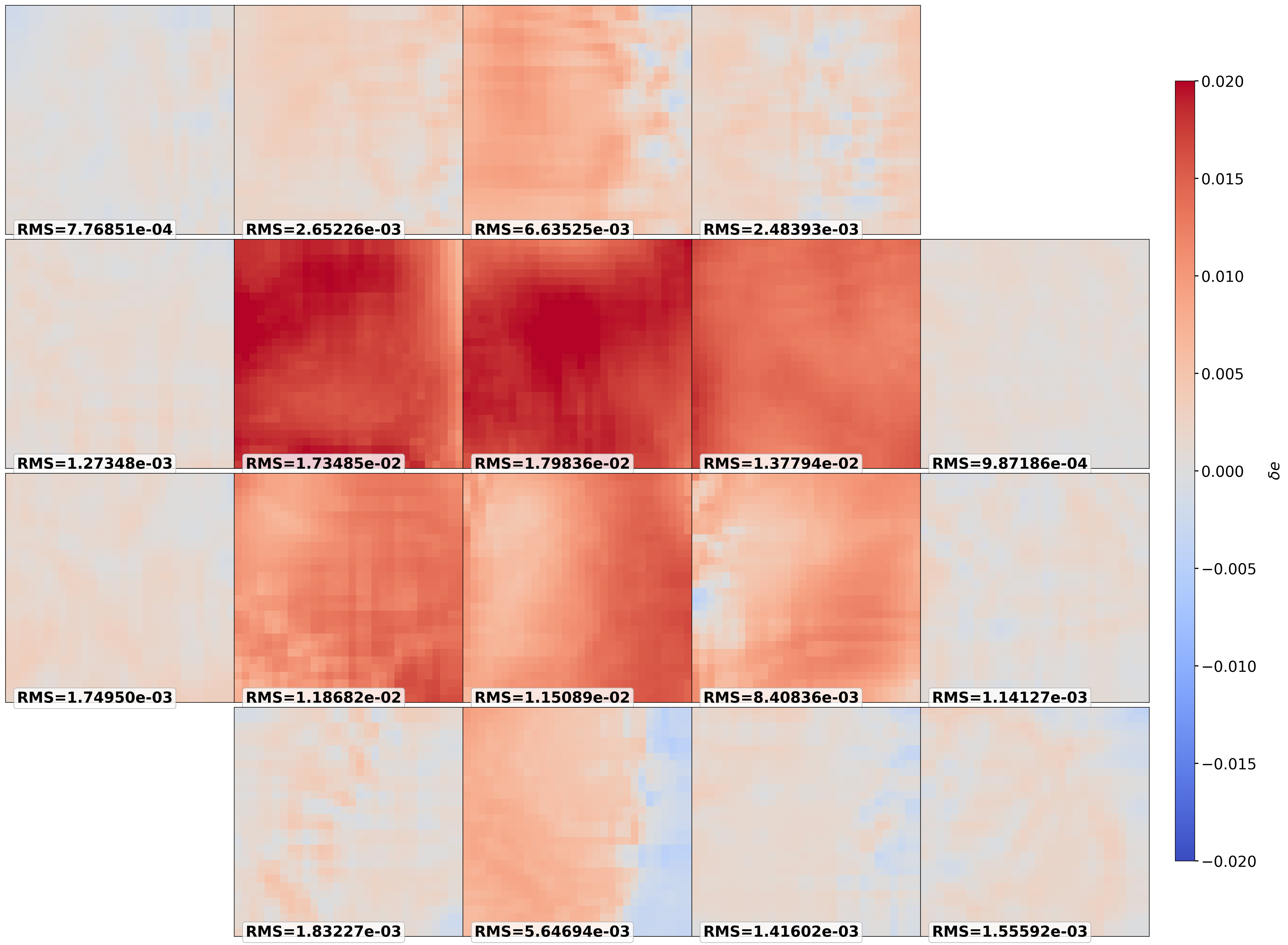}
    \caption{$\delta e$ distribution of \texttt{DATASET 3}}
    \label{Model.Residual.e.blur}
  \end{subfigure}
  \caption{Residual distributions of PSF size ($\delta R^2/R^2$, panel a) and ellipticity ($\delta e$, panel b) across all 18 CCDs for  \texttt{CSST-PSFNet}, evaluated on \texttt{DATASET 3}.}
  \label{fig:Model.Residual_R_e.blur}
\end{figure}

\begin{figure}[h]
  \centering
  \begin{subfigure}{0.48\textwidth}
    \includegraphics[width=\textwidth]{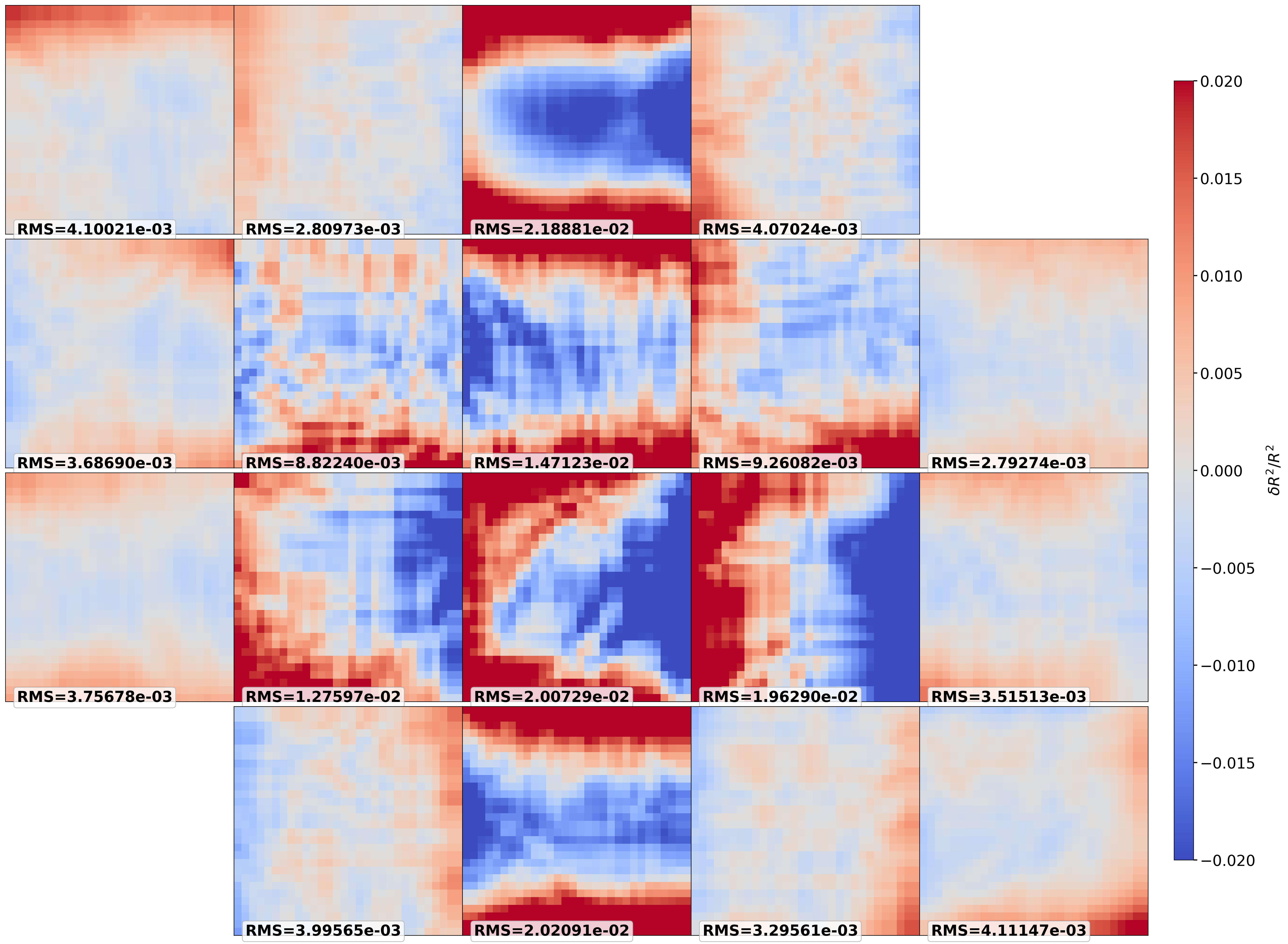}
    \caption{$\delta R^2/R^2$ distribution of \texttt{DATASET 3}}
    \label{PSFEx.Residual.R.blur}
  \end{subfigure}
  \hfill
  \begin{subfigure}{0.48\textwidth}
    \includegraphics[width=\textwidth]{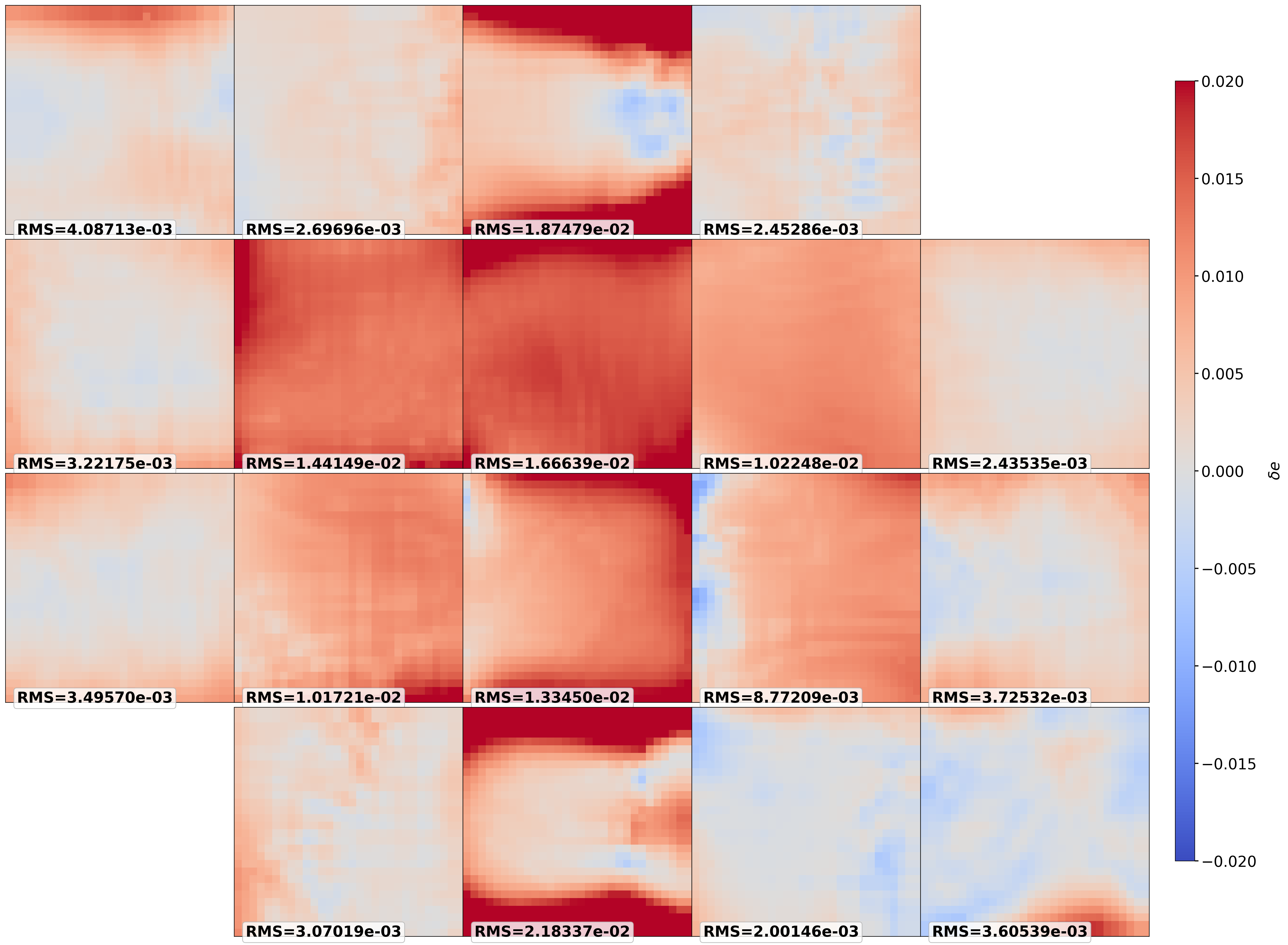}
    \caption{$\delta e$ distribution of \texttt{DATASET 3}}
    \label{PSFEx.Residual.e.blur}
  \end{subfigure}
  \caption{
    Residual distributions of PSF size ($\delta R^2/R^2$, panel a) and ellipticity ($\delta e$, panel b) across all 18 CCDs for \texttt{PSFEx}, evaluated on \texttt{DATASET 3}. 
    }
  \label{fig:PSFEx.Residual_R_e.blur}
\end{figure}

Figures~\ref{fig:Model.Residual_R_e.blur} and~\ref{fig:PSFEx.Residual_R_e.blur} 
present the spatial distributions of the size and ellipticity residuals, 
$\delta R^2/R^2$ and $\delta e$. 
\texttt{DATASET 3} effectively broadens the nominal PSFs, reducing the degree of undersampling and thus mitigating high-frequency discrepancies between the predicted and true PSFs.
Nevertheless, \texttt{PSFEx} still exhibits noticeable artificial structures, particularly along the edges of the $NUV, u, g$ bands, revealing its inherent limitations in interpolation-based modeling. In contrast, \texttt{CSST-PSFNet} achieves consistently smaller residuals across all bands and shows no visible artifacts, with the improvement being most pronounced in the severely undersampled blue bands. These results confirm that the proposed model provides more accurate and physically stable PSF reconstructions under on-orbit–like conditions, demonstrating superior robustness and generalization compared to the traditional interpolation approach.

\subsection{Cross-band Consistency and Robustness}
\label{sec: Cross-band Consistency and Robustness}
\begin{figure}[H]
  \centering
  \begin{subfigure}{0.45\textwidth}
    \includegraphics[width=\textwidth]{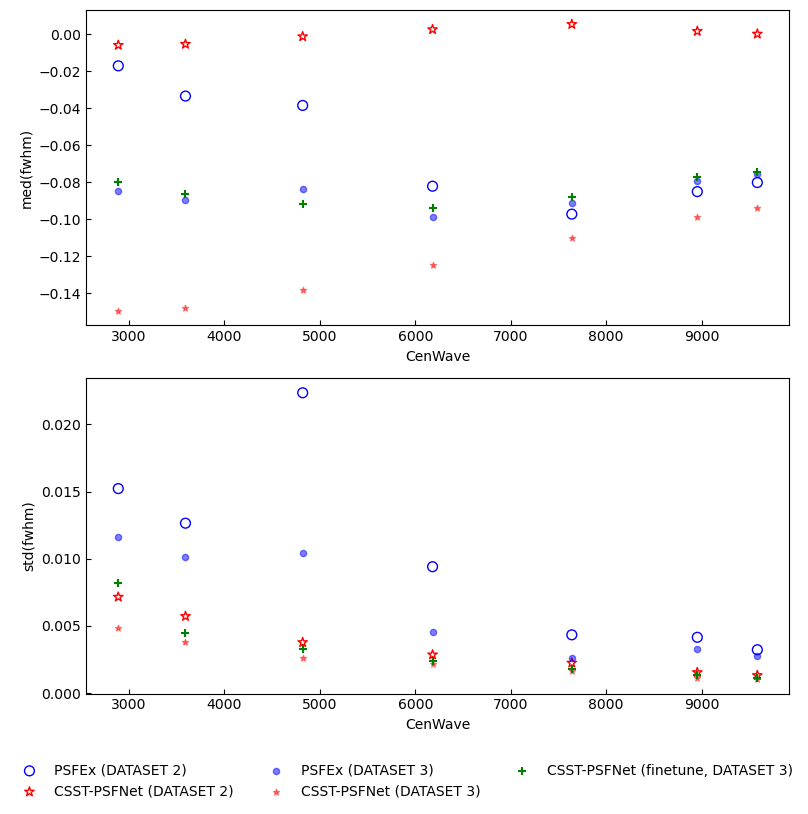}
    \label{cenwave_fwhm_sigma}
  \end{subfigure}
  \hfill
  \begin{subfigure}{0.45\textwidth}
    \includegraphics[width=\textwidth]{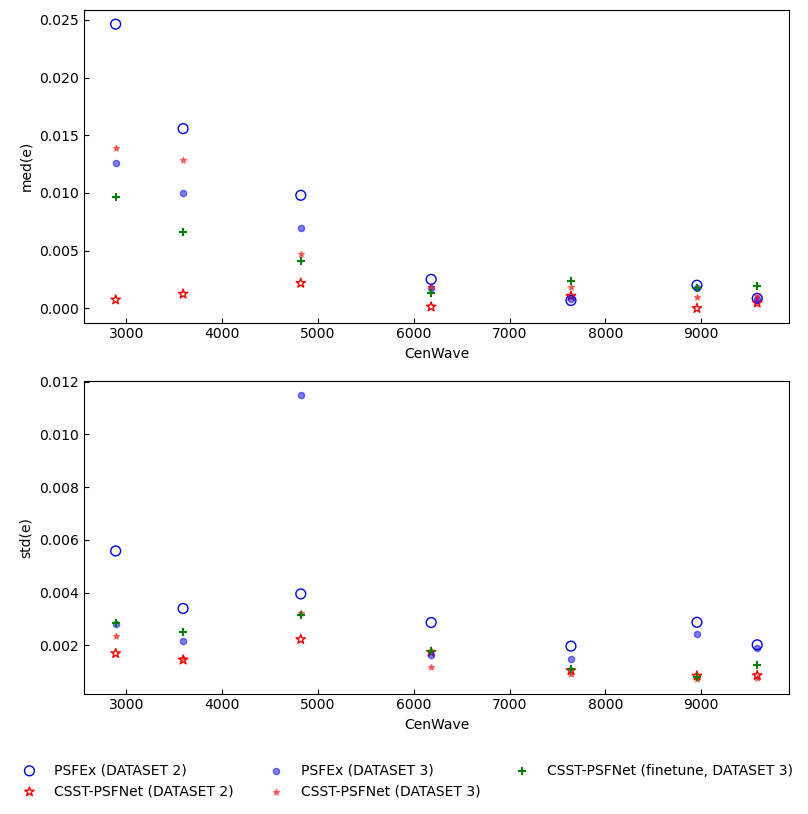}
    \label{cenwave_e_sigma}
  \end{subfigure}
  \caption{Cross-band statistics of PSF reconstruction accuracy grouped by filter central wavelength.
  The \textbf{top panels} show the median residuals of the size ratio $\delta R^2/R^2$ and the ellipticity components $\delta e$ across all CCDs in each band.
  The \textbf{bottom panels} present the corresponding dispersions $\sigma({\delta R^2/R^2})$ and $\sigma({\delta e})$, evaluating the stability of the reconstruction.
  Blue markers denote results from \texttt{PSFEx}, while red markers denote results from \texttt{CSST-PSFNet}.
  For \texttt{DATASET 2}, results are shown as large open circles (PSFEx) and large open star markers (CSST-PSFNet).
  For the Gaussian-blurred test set \texttt{DATASET 3}, results are shown as small filled circles (PSFEx) and small filled star markers (CSST-PSFNet).
  Green plus symbols indicate the performance of \texttt{CSST-PSFNet} on \texttt{DATASET 3} after fine-tuning with \texttt{PSFEx}-predicted PSFs as weak labels (see Section~\ref{sec: discussion}).}
  \label{fig:med_std_cenwave}
\end{figure}

Figure~\ref{fig:med_std_cenwave} summarizes the cross-band residual statistics of PSF size and ellipticity for both \texttt{PSFEx} and \texttt{CSST-PSFNet} under \texttt{DATASET 2} and \texttt{DATASET 3}.
Across all photometric bands, \texttt{CSST-PSFNet} exhibits dispersions substantially smaller than those of \texttt{PSFEx}, demonstrating superior cross-band consistency. In terms of PSF size, the dispersion $\sigma({\delta R^2/R^2})$ for \texttt{CSST-PSFNet} remains below 0.005 in the focal plane and drops to $\sim0.002$ in the redder bands, while \texttt{PSFEx} shows significantly higher values, particularly in the blue filters where undersampling is most severe.

After Gaussian broadening, both methods show reduced $\sigma({\delta R^2/R^2})$, reflecting the smoothing effect of isotropic convolution on high-frequency structures. For ellipticity, \texttt{PSFEx} shows a uniform decrease in $\sigma({\delta e})$ across all bands, while \texttt{CSST-PSFNet} exhibits a modest increase in the bluest filters and remains nearly unchanged toward the red. This trend is consistent with the physical impact of Gaussian blurring, which weakens the anisotropic features of sharp blue band PSFs and thus reduces the morphological cues available to the network.

Despite these effects, \texttt{CSST-PSFNet} maintains lower residual dispersions in both $\sigma({\delta R^2/R^2})$ and $\sigma({\delta e})$ across all bands and conditions, indicating that the model is more robust in recovering PSF size and anisotropy even under degraded or smoothed observational scenarios.

\section{Discussions}
\label{sec: discussion}
\subsection{Weak Supervision, Model Limitations, and Future Calibration}
\label{sec: discussion1}
This study evaluates the performance of \texttt{CSST-PSFNet} using high-fidelity simulations generated by the CSST Main Survey Simulator \citep{wei_2025, Bai_2025}. 
While these simulations incorporate realistic optical modeling, focal-plane geometry, and detector physics, they inevitably represent an idealized approximation of on-orbit conditions.
Real observations may exhibit complex instrument-state evolution, time-dependent perturbations (thermal drift, micro-vibrations, detector aging), and unforeseen optical artifacts not fully captured in pre-launch simulations. These limitations necessitate a strategy for model adaptation using real on-orbit data.

To explore adaptation under realistic constraints, we conducted a fine-tuning experiment (Section~\ref{sec: Cross-band Consistency and Robustness}, Figure~\ref{fig:med_std_cenwave}) in which PSFEx-estimated PSFs serve as weak supervisory labels. After fine-tuning, the residuals in median $\delta R^2/R^2$ (particularly prominent in blue bands) reduce toward PSFEx levels while maintaining lower random scatter. 
This result indicates that \texttt{CSST-PSFNet} can align with PSFEx-level reconstructions when simulation-based priors differ from alternative PSF modeling approaches.
The experiment demonstrates consistency under weak supervision, rather than direct correction of unknown residuals in the true on-orbit PSF.

The experiment reveals three fundamental characteristics. 
First, when trained on labels carrying residuals (PSFEx's polynomial limitations), these residuals necessarily propagate into the neural network, establishing a performance ceiling bounded by the supervision source. 
Second, simulation-trained representations provide regularization that stabilizes fine-tuning under noisy supervision, maintaining lower scatter than PSFEx despite noisy labels. 
Third, any retained features differing from PSFEx could represent either genuine physics (diffraction patterns, inter-CCD correlations) or simulation artifacts.
Fine-tuning provides a robust fallback when simulations prove inaccurate, matching traditional methods in their optimal regime.

To surpass this accuracy ceiling, a possible future direction is to adopt an iterative refinement strategy in which simulator parameters are progressively updated using real on-orbit observations. Recent advances in generative modeling have demonstrated the effectiveness of such iterative calibration approaches. For example, \citet{Bai2024} introduced EMDiffusion, an expectation–maximization framework for training diffusion models from corrupted data, while \citet{Barco2025} and \citet{Hosseintabar2025} explored related iterative updating schemes with improved convergence behavior under misspecified priors.
Adapting these ideas to the CSST context would involve treating simulator parameters as learnable quantities that are iteratively refined using residuals between reconstructed PSFs and observed stellar images. In a hierarchical calibration setting, the neural network parameters and simulator parameters could be alternately updated as additional on-orbit data accumulate.
Such an approach may provide a pathway toward alleviating the weak-supervision performance ceiling by using real observations as the ultimate constraint rather than intermediate estimators. While beyond the scope of the present work, this direction represents a natural extension of the framework once sufficient on-orbit data becomes available.

\subsection{Operational Advantages and Broader Implications}
While \texttt{CSST-PSFNet} is developed for the CSST imaging survey, several design features make the framework broadly relevant to other space- and ground-based telescopes. The model conditions on focal-plane coordinates and detector identifiers, enabling it to represent both smooth spatial PSF variations and inter-detector discontinuities. This structural flexibility is compatible with a wide range of instruments, including multi-CCD wide-field imagers (e.g., LSST, DECam, Hyper Suprime-Cam) and undersampled space telescopes such as Euclid VIS, Roman WFI, and JWST/NIRCam, where accurate PSF reconstruction is critical for weak-lensing and photometric precision.

However, PSF morphology is fundamentally determined by a telescope’s specific optical and detector design. Consequently, while the architecture of \texttt{CSST-PSFNet} is transferable, the learned parameters are not. Adapting the model to a new instrument would therefore require retraining with that instrument’s own PSF data—either from detailed optical simulations or from on-sky calibration stars. In practice, this adaptation would mainly involve (i) replacing the CSST-specific coordinate and detector embeddings with those reflecting the target focal-plane geometry, and (ii) re-training the network on a modest set of instrument-specific PSFs. The adaptation experiment in Section~\ref{sec: discussion1}, in which the model is updated using weak labels derived from \texttt{PSFEx}, demonstrates that \texttt{CSST-PSFNet} can integrate approximate PSF information effectively, suggesting that only limited calibration data might be necessary for transfer to other instruments.

Looking ahead, data-driven models such as \texttt{CSST-PSFNet} may play an increasingly important role in future instrument calibration pipelines. Traditional calibration strategies, relying on polynomial interpolation, sparse calibration stars, or static optical models, may become insufficient as surveys demand higher resolution, tighter stability, and stricter control of weak-lensing residuals. A generative PSF model that can merge simulation data, laboratory measurements, and on-orbit observations offers a more flexible alternative.

In a future calibration architecture, a deep PSF model could serve as a continuously updated representation of the instrument’s optical state. The network could be initialized with pre-launch simulations and refined progressively using commissioning data, approximate PSF estimates, or dedicated calibration fields. The conditioning strategy further enables the inclusion of temporal or environmental metadata—such as orbital temperature, alignment sensors, or attitude-control diagnostics—thereby potentially enabling the construction of time-dependent PSF models. 

\section{Conclusions}
\label{sec: conclusion}
In this work, we introduced \texttt{CSST-PSFNet}, a deep-learning framework for high-fidelity PSF reconstruction, specifically designed for the CSST. By integrating convolutional feature extraction, Transformer-based global aggregation, and variational latent modeling, the network effectively captures both local PSF structures and large-scale variations across the multi-CCD focal plane. The incorporation of geometric coordinate encoding and CCD-specific embeddings further enables the model to represent spatially varying, detector-dependent PSF behavior in a unified manner.

Using high-resolution simulated star–PSF pairs generated by the CSST Main Survey Simulator, \texttt{CSST-PSFNet} achieves improved pixel-level accuracy and more precise recovery of weak-lensing–sensitive shape parameters compared to the widely used \texttt{PSFEx} algorithm under the simulated conditions considered in this study. The model further demonstrates stable performance within the tested Gaussian-blur perturbation regime, indicating robustness to moderate, controlled deviations from the nominal PSF morphology. 

To assess the applicability of the model under realistic on-orbit conditions, we further performed an adaptation test in which the ground truth PSF is assumed unavailable, and PSFs estimated by \texttt{PSFEx} serve as weak supervision. Through limited fine-tuning, the model aligns with PSFEx-level reconstruction accuracy while maintaining reduced statistical scatter. This result demonstrates consistency under weak supervision rather than direct correction of unknown residuals, and suggests that the framework can flexibly incorporate approximate PSF information when high-fidelity supervision is not accessible.

We emphasize that the present validation is performed within a high-fidelity simulation regime. Full assessment under operational conditions will require real on-orbit observations, including multi-epoch calibration stars and time-dependent instrument monitoring. With such data, hierarchical or iterative calibration strategies may further refine the physical consistency of the model. Within its current scope, \texttt{CSST-PSFNet} provides a scalable and instrument-aware approach to PSF reconstruction for undersampled, wide-field space imaging surveys such as CSST.

\section*{Data Availability}
The simulated datasets used in this study were generated using the CSST Main Survey Simulator (csst\_msc\_sim) within the CSST collaboration framework. Due to the large data volume and the dependence on internal simulation components and configurations of the CSST project, the processed datasets and trained model weights cannot be publicly deposited at present. The core model implementation and data-processing framework are publicly available via the authors’ GitHub repository. However, full reproduction of the datasets requires access to internal CSST simulation resources and configurations that are not currently publicly accessible. The simulation configurations and methodological details described in Section~2 document the data-generation procedure and workflow. Access to the processed datasets and trained model weights may be granted by the corresponding author upon reasonable request and subject to collaboration approval.

\begin{acknowledgments}
This work is based in part on the software created by the CSST Simulation Team, which is supported by the CSST scientific data processing and analysis system of the China Manned Space Project. The authors acknowledge support by the CSST Scientific Data Processing System (No. CMS-CSST-2025-A11), the China National Science Foundation (NSFC) (Nos. 12433012, 12373097), the China Manned Space Project (grant Nos. CMS-CSST-2025-A19, CMS-CSST-2025-A21), and the National Astronomical Observatories, Chinese Academy of Sciences (Project No. E4ZR050901).
\end{acknowledgments}

%% For this sample we use BibTeX plus aasjournalv7.bst to generate the

\bibliography{sample701}{}

@article{Euclid_2025,
   title={Euclid: I. Overview of the Euclid mission},
   volume={697},
   ISSN={1432-0746},
   url={http://dx.doi.org/10.1051/0004-6361/202450810},
   DOI={10.1051/0004-6361/202450810},
   journal={Astronomy \& Astrophysics},
   publisher={EDP Sciences},
   author={Mellier, Y. and Abdurro’uf and Acevedo Barroso, J. A. and Achúcarro, A. and Adamek, J. and Adam, R. and Addison, G. E. and Aghanim, N. and Aguena, M. and Ajani, V. and Akrami, Y. and Al-Bahlawan, A. and Alavi, A. and Albuquerque, I. S. and Alestas, G. and Alguero, G. and Allaoui, A. and Allen, S. W. and Allevato, V. and Alonso-Tetilla, A. V. and Altieri, B. and Alvarez-Candal, A. and Alvi, S. and Amara, A. and Amendola, L. and Amiaux, J. and Andika, I. T. and Andreon, S. and Andrews, A. and Angora, G. and Angulo, R. E. and Annibali, F. and Anselmi, A. and Anselmi, S. and Arcari, S. and Archidiacono, M. and Aricò, G. and Arnaud, M. and Arnouts, S. and Asgari, M. and Asorey, J. and Atayde, L. and Atek, H. and Atrio-Barandela, F. and Aubert, M. and Aubourg, E. and Auphan, T. and Auricchio, N. and Aussel, B. and Aussel, H. and Avelino, P. P. and Avgoustidis, A. and Avila, S. and Awan, S. and Azzollini, R. and Baccigalupi, C. and Bachelet, E. and Bacon, D. and Baes, M. and Bagley, M. B. and Bahr-Kalus, B. and Balaguera-Antolinez, A. and Balbinot, E. and Balcells, M. and Baldi, M. and Baldry, I. and Balestra, A. and Ballardini, M. and Ballester, O. and Balogh, M. and Bañados, E. and Barbier, R. and Bardelli, S. and Baron, M. and Barreiro, T. and Barrena, R. and Barriere, J.-C. and Barros, B. J. and Barthelemy, A. and Bartolo, N. and Basset, A. and Battaglia, P. and Battisti, A. J. and Baugh, C. M. and Baumont, L. and Bazzanini, L. and Beaulieu, J.-P. and Beckmann, V. and Belikov, A. N. and Bel, J. and Bellagamba, F. and Bella, M. and Bellini, E. and Benabed, K. and Bender, R. and Benevento, G. and Bennett, C. L. and Benson, K. and Bergamini, P. and Bermejo-Climent, J. R. and Bernardeau, F. and Bertacca, D. and Berthe, M. and Berthier, J. and Bethermin, M. and Beutler, F. and Bevillon, C. and Bhargava, S. and Bhatawdekar, R. and Bianchi, D. and Bisigello, L. and Biviano, A. and Blake, R. P. and Blanchard, A. and Blazek, J. and Blot, L. and Bosco, A. and Bodendorf, C. and Boenke, T. and Böhringer, H. and Boldrini, P. and Bolzonella, M. and Bonchi, A. and Bonici, M. and Bonino, D. and Bonino, L. and Bonvin, C. and Bon, W. and Booth, J. T. and Borgani, S. and Borlaff, A. S. and Borsato, E. and Bosco, A. and Bose, B. and Botticella, M. T. and Boucaud, A. and Bouche, F. and Boucher, J. S. and Boutigny, D. and Bouvard, T. and Bouwens, R. and Bouy, H. and Bowler, R. A. A. and Bozza, V. and Bozzo, E. and Branchini, E. and Brando, G. and Brau-Nogue, S. and Brekke, P. and Bremer, M. N. and Brescia, M. and Breton, M.-A. and Brinchmann, J. and Brinckmann, T. and Brockley-Blatt, C. and Brodwin, M. and Brouard, L. and Brown, M. L. and Bruton, S. and Bucko, J. and Buddelmeijer, H. and Buenadicha, G. and Buitrago, F. and Burger, P. and Burigana, C. and Busillo, V. and Busonero, D. and Cabanac, R. and Cabayol-Garcia, L. and Cagliari, M. S. and Caillat, A. and Caillat, L. and Calabrese, M. and Calabro, A. and Calderone, G. and Calura, F. and Camacho Quevedo, B. and Camera, S. and Campos, L. and Cañas-Herrera, G. and Candini, G. P. and Cantiello, M. and Capobianco, V. and Cappellaro, E. and Cappelluti, N. and Cappi, A. and Caputi, K. I. and Cara, C. and Carbone, C. and Cardone, V. F. and Carella, E. and Carlberg, R. G. and Carle, M. and Carminati, L. and Caro, F. and Carrasco, J. M. and Carretero, J. and Carrilho, P. and Carron Duque, J. and Carry, B. and Carvalho, A. and Carvalho, C. S. and Casas, R. and Casas, S. and Casenove, P. and Casey, C. M. and Cassata, P. and Castander, F. J. and Castelao, D. and Castellano, M. and Castiblanco, L. and Castignani, G. and Castro, T. and Cavet, C. and Cavuoti, S. and Chabaud, P.-Y. and Chambers, K. C. and Charles, Y. and Charlot, S. and Chartab, N. and Chary, R. and Chaumeil, F. and Cho, H. and Chon, G. and Ciancetta, E. and Ciliegi, P. and Cimatti, A. and Cimino, M. and Cioni, M.-R. L. and Claydon, R. and Cleland, C. and Clément, B. and Clements, D. L. and Clerc, N. and Clesse, S. and Codis, S. and Cogato, F. and Colbert, J. and Cole, R. E. and Coles, P. and Collett, T. E. and Collins, R. S. and Colodro-Conde, C. and Colombo, C. and Combes, F. and Conforti, V. and Congedo, G. and Conseil, S. and Conselice, C. J. and Contarini, S. and Contini, T. and Conversi, L. and Cooray, A. R. and Copin, Y. and Corasaniti, P.-S. and Corcho-Caballero, P. and Corcione, L. and Cordes, O. and Corpace, O. and Correnti, M. and Costanzi, M. and Costille, A. and Courbin, F. and Courcoult Mifsud, L. and Courtois, H. M. and Cousinou, M.-C. and Covone, G. and Cowell, T. and Cragg, C. and Cresci, G. and Cristiani, S. and Crocce, M. and Cropper, M. and Crouzet, P. E. and Csizi, B. and Cuby, J.-G. and Cucchetti, E. and Cucciati, O. and Cuillandre, J.-C. and Cunha, P. A. C. and Cuozzo, V. and Daddi, E. and D’Addona, M. and Dafonte, C. and Dagoneau, N. and Dalessandro, E. and Dalton, G. B. and D’Amico, G. and Dannerbauer, H. and Danto, P. and Das, I. and Da Silva, A. and da Silva, R. and d’Assignies Doumerg, W. and Daste, G. and Davies, J. E. and Davini, S. and Dayal, P. and de Boer, T. and Decarli, R. and De Caro, B. and Degaudenzi, H. and Degni, G. and de Jong, J. T. A. and de la Bella, L. F. and de la Torre, S. and Delhaise, F. and Delley, D. and Delucchi, G. and De Lucia, G. and Denniston, J. and De Paolis, F. and De Petris, M. and Derosa, A. and Desai, S. and Desjacques, V. and Despali, G. and Desprez, G. and De Vicente-Albendea, J. and Deville, Y. and Dias, J. D. F. and Díaz-Sánchez, A. and Diaz, J. J. and Di Domizio, S. and Diego, J. M. and Di Ferdinando, D. and Di Giorgio, A. M. and Dimauro, P. and Dinis, J. and Dolag, K. and Dolding, C. and Dole, H. and Domínguez Sánchez, H. and Doré, O. and Dournac, F. and Douspis, M. and Dreihahn, H. and Droge, B. and Dryer, B. and Dubath, F. and Duc, P.-A. and Ducret, F. and Duffy, C. and Dufresne, F. and Duncan, C. A. J. and Dupac, X. and Duret, V. and Durrer, R. and Durret, F. and Dusini, S. and Ealet, A. and Eggemeier, A. and Eisenhardt, P. R. M. and Elbaz, D. and Elkhashab, M. Y. and Ellien, A. and Endicott, J. and Enia, A. and Erben, T. and Escartin Vigo, J. A. and Escoffier, S. and Escudero Sanz, I. and Essert, J. and Ettori, S. and Ezziati, M. and Fabbian, G. and Fabricius, M. and Fang, Y. and Farina, A. and Farina, M. and Farinelli, R. and Farrens, S. and Faustini, F. and Feltre, A. and Ferguson, A. M. N. and Ferrando, P. and Ferrari, A. G. and Ferré-Mateu, A. and Ferreira, P. G. and Ferreras, I. and Ferrero, I. and Ferriol, S. and Ferruit, P. and Filleul, D. and Finelli, F. and Finkelstein, S. L. and Finoguenov, A. and Fiorini, B. and Flentge, F. and Focardi, P. and Fonseca, J. and Fontana, A. and Fontanot, F. and Fornari, F. and Fosalba, P. and Fossati, M. and Fotopoulou, S. and Fouchez, D. and Fourmanoit, N. and Frailis, M. and Fraix-Burnet, D. and Franceschi, E. and Franco, A. and Franzetti, P. and Freihoefer, J. and Frenk, C. S. and Frittoli, G. and Frugier, P.-A. and Frusciante, N. and Fumagalli, A. and Fumagalli, M. and Fumana, M. and Fu, Y. and Gabarra, L. and Galeotta, S. and Galluccio, L. and Ganga, K. and Gao, H. and García-Bellido, J. and Garcia, K. and Gardner, J. P. and Garilli, B. and Gaspar-Venancio, L.-M. and Gasparetto, T. and Gautard, V. and Gavazzi, R. and Gaztanaga, E. and Genolet, L. and Genova Santos, R. and Gentile, F. and George, K. and Gerbino, M. and Ghaffari, Z. and Giacomini, F. and Gianotti, F. and Gibb, G. P. S. and Gillard, W. and Gillis, B. and Ginolfi, M. and Giocoli, C. and Girardi, M. and Giri, S. K. and Goh, L. W. K. and Gómez-Alvarez, P. and Gonzalez-Perez, V. and Gonzalez, A. H. and Gonzalez, E. J. and Gonzalez, J. C. and Gouyou Beauchamps, S. and Gozaliasl, G. and Gracia-Carpio, J. and Grandis, S. and Granett, B. R. and Granvik, M. and Grazian, A. and Gregorio, A. and Grenet, C. and Grillo, C. and Grupp, F. and Gruppioni, C. and Gruppuso, A. and Guerbuez, C. and Guerrini, S. and Guidi, M. and Guillard, P. and Gutierrez, C. M. and Guttridge, P. and Guzzo, L. and Gwyn, S. and Haapala, J. and Haase, J. and Haddow, C. R. and Hailey, M. and Hall, A. and Hall, D. and Hamaus, N. and Haridasu, B. S. and Harnois-Déraps, J. and Harper, C. and Hartley, W. G. and Hasinger, G. and Hassani, F. and Hatch, N. A. and Haugan, S. V. H. and Häußler, B. and Heavens, A. and Heisenberg, L. and Helmi, A. and Helou, G. and Hemmati, S. and Henares, K. and Herent, O. and Hernández-Monteagudo, C. and Heuberger, T. and Hewett, P. C. and Heydenreich, S. and Hildebrandt, H. and Hirschmann, M. and Hjorth, J. and Hoar, J. and Hoekstra, H. and Holland, A. D. and Holliman, M. S. and Holmes, W. and Hook, I. and Horeau, B. and Hormuth, F. and Hornstrup, A. and Hosseini, S. and Hu, D. and Hudelot, P. and Hudson, M. J. and Huertas-Company, M. and Huff, E. M. and Hughes, A. C. N. and Humphrey, A. and Hunt, L. K. and Huynh, D. D. and Ibata, R. and Ichikawa, K. and Iglesias-Groth, S. and Ilbert, O. and Ilić, S. and Ingoglia, L. and Iodice, E. and Israel, H. and Israelsson, U. E. and Izzo, L. and Jablonka, P. and Jackson, N. and Jacobson, J. and Jafariyazani, M. and Jahnke, K. and Jain, B. and Jansen, H. and Jarvis, M. J. and Jasche, J. and Jauzac, M. and Jeffrey, N. and Jhabvala, M. and Jimenez-Teja, Y. and Jimenez Muñoz, A. and Joachimi, B. and Johansson, P. H. and Joudaki, S. and Jullo, E. and Kajava, J. J. E. and Kang, Y. and Kannawadi, A. and Kansal, V. and Karagiannis, D. and Kärcher, M. and Kashlinsky, A. and Kazandjian, M. V. and Keck, F. and Keihänen, E. and Kerins, E. and Kermiche, S. and Khalil, A. and Kiessling, A. and Kiiveri, K. and Kilbinger, M. and Kim, J. and King, R. and Kirkpatrick, C. C. and Kitching, T. and Kluge, M. and Knabenhans, M. and Knapen, J. H. and Knebe, A. and Kneib, J.-P. and Kohley, R. and Koopmans, L. V. E. and Koskinen, H. and Koulouridis, E. and Kou, R. and Kovács, A. and Kovačić, I. and Kowalczyk, A. and Koyama, K. and Kraljic, K. and Krause, O. and Kruk, S. and Kubik, B. and Kuchner, U. and Kuijken, K. and Kümmel, M. and Kunz, M. and Kurki-Suonio, H. and Lacasa, F. and Lacey, C. G. and La Franca, F. and Lagarde, N. and Lahav, O. and Laigle, C. and La Marca, A. and La Marle, O. and Lamine, B. and Lam, M. C. and Lançon, A. and Landt, H. and Langer, M. and Lapi, A. and Larcheveque, C. and Larsen, S. S. and Lattanzi, M. and Laudisio, F. and Laugier, D. and Laureijs, R. and Laurent, V. and Lavaux, G. and Lawrenson, A. and Lazanu, A. and Lazeyras, T. and Le Boulc’h, Q. and Le Brun, A. M. C. and Le Brun, V. and Leclercq, F. and Lee, S. and Le Graet, J. and Legrand, L. and Leirvik, K. N. and Le Jeune, M. and Lembo, M. and Le Mignant, D. and Lepinzan, M. D. and Lepori, F. and Le Reun, A. and Leroy, G. and Lesci, G. F. and Lesgourgues, J. and Leuzzi, L. and Levi, M. E. and Liaudat, T. I. and Libet, G. and Liebing, P. and Ligori, S. and Lilje, P. B. and Lin, C.-C. and Linde, D. and Linder, E. and Lindholm, V. and Linke, L. and Li, S.-S. and Liu, S. J. and Lloro, I. and Lobo, F. S. N. and Lodieu, N. and Lombardi, M. and Lombriser, L. and Lonare, P. and Longo, G. and López-Caniego, M. and Lopez Lopez, X. and Lorenzo Alvarez, J. and Loureiro, A. and Loveday, J. and Lusso, E. and Macias-Perez, J. and Maciaszek, T. and Maggio, G. and Magliocchetti, M. and Magnard, F. and Magnier, E. A. and Magro, A. and Mahler, G. and Mainetti, G. and Maino, D. and Maiorano, E. and Maiorano, E. and Malavasi, N. and Mamon, G. A. and Mancini, C. and Mandelbaum, R. and Manera, M. and Manjón-García, A. and Mannucci, F. and Mansutti, O. and Manteiga Outeiro, M. and Maoli, R. and Maraston, C. and Marcin, S. and Marcos-Arenal, P. and Margalef-Bentabol, B. and Marggraf, O. and Marinucci, D. and Marinucci, M. and Markovic, K. and Marleau, F. R. and Marpaud, J. and Martignac, J. and Martín-Fleitas, J. and Martin-Moruno, P. and Martin, E. L. and Martinelli, M. and Martinet, N. and Martin, H. and Martins, C. J. A. P. and Marulli, F. and Massari, D. and Massey, R. and Masters, D. C. and Matarrese, S. and Matsuoka, Y. and Matthew, S. and Maughan, B. J. and Mauri, N. and Maurin, L. and Maurogordato, S. and McCarthy, K. and McConnachie, A. W. and McCracken, H. J. and McDonald, I. and McEwen, J. D. and McPartland, C. J. R. and Medinaceli, E. and Mehta, V. and Mei, S. and Melchior, M. and Melin, J.-B. and Ménard, B. and Mendes, J. and Mendez-Abreu, J. and Meneghetti, M. and Mercurio, A. and Merlin, E. and Metcalf, R. B. and Meylan, G. and Migliaccio, M. and Mignoli, M. and Miller, L. and Miluzio, M. and Milvang-Jensen, B. and Mimoso, J. P. and Miquel, R. and Miyatake, H. and Mobasher, B. and Mohr, J. J. and Monaco, P. and Monguió, M. and Montoro, A. and Mora, A. and Moradinezhad Dizgah, A. and Moresco, M. and Moretti, C. and Morgante, G. and Morisset, N. and Moriya, T. J. and Morris, P. W. and Mortlock, D. J. and Moscardini, L. and Mota, D. F. and Mottet, S. and Moustakas, L. A. and Moutard, T. and Müller, T. and Munari, E. and Murphree, G. and Murray, C. and Murray, N. and Musi, P. and Nadathur, S. and Nagam, B. C. and Nagao, T. and Naidoo, K. and Nakajima, R. and Nally, C. and Natoli, P. and Navarro-Alsina, A. and Navarro Girones, D. and Neissner, C. and Nersesian, A. and Nesseris, S. and Nguyen-Kim, H. N. and Nicastro, L. and Nichol, R. C. and Nielbock, M. and Niemi, S.-M. and Nieto, S. and Nilsson, K. and Noller, J. and Norberg, P. and Nouri-Zonoz, A. and Ntelis, P. and Nucita, A. A. and Nugent, P. and Nunes, N. J. and Nutma, T. and Ocampo, I. and Odier, J. and Oesch, P. A. and Oguri, M. and Magalhaes Oliveira, D. and Onoue, M. and Oosterbroek, T. and Oppizzi, F. and Ordenovic, C. and Osato, K. and Pacaud, F. and Pace, F. and Padilla, C. and Paech, K. and Pagano, L. and Page, M. J. and Palazzi, E. and Paltani, S. and Pamuk, S. and Pandolfi, S. and Paoletti, D. and Paolillo, M. and Papaderos, P. and Pardede, K. and Parimbelli, G. and Parmar, A. and Partmann, C. and Pasian, F. and Passalacqua, F. and Paterson, K. and Patrizii, L. and Pattison, C. and Paulino-Afonso, A. and Paviot, R. and Peacock, J. A. and Pearce, F. R. and Pedersen, K. and Peel, A. and Peletier, R. F. and Pellejero Ibanez, M. and Pello, R. and Penny, M. T. and Percival, W. J. and Perez-Garrido, A. and Perotto, L. and Pettorino, V. and Pezzotta, A. and Pezzuto, S. and Philippon, A. and Pierre, M. and Piersanti, O. and Pietroni, M. and Piga, L. and Pilo, L. and Pires, S. and Pisani, A. and Pizzella, A. and Pizzuti, L. and Plana, C. and Polenta, G. and Pollack, J. E. and Poncet, M. and Pöntinen, M. and Pool, P. and Popa, L. A. and Popa, V. and Popp, J. and Porciani, C. and Porth, L. and Potter, D. and Poulain, M. and Pourtsidou, A. and Pozzetti, L. and Prandoni, I. and Pratt, G. W. and Prezelus, S. and Prieto, E. and Pugno, A. and Quai, S. and Quilley, L. and Racca, G. D. and Raccanelli, A. and Rácz, G. and Radinović, S. and Radovich, M. and Ragagnin, A. and Ragnit, U. and Raison, F. and Ramos-Chernenko, N. and Ranc, C. and Rasera, Y. and Raylet, N. and Rebolo, R. and Refregier, A. and Reimberg, P. and Reiprich, T. H. and Renk, F. and Renzi, A. and Retre, J. and Revaz, Y. and Reylé, C. and Reynolds, L. and Rhodes, J. and Ricci, F. and Ricci, M. and Riccio, G. and Ricken, S. O. and Rissanen, S. and Risso, I. and Rix, H.-W. and Robin, A. C. and Rocca-Volmerange, B. and Rocci, P.-F. and Rodenhuis, M. and Rodighiero, G. and Rodriguez Monroy, M. and Rollins, R. P. and Romanello, M. and Roman, J. and Romelli, E. and Romero-Gomez, M. and Roncarelli, M. and Rosati, P. and Rosset, C. and Rossetti, E. and Roster, W. and Rottgering, H. J. A. and Rozas-Fernández, A. and Ruane, K. and Rubino-Martin, J. A. and Rudolph, A. and Ruppin, F. and Rusholme, B. and Sacquegna, S. and Sáez-Casares, I. and Saga, S. and Saglia, R. and Sahlén, M. and Saifollahi, T. and Sakr, Z. and Salvalaggio, J. and Salvaterra, R. and Salvati, L. and Salvato, M. and Salvignol, J.-C. and Sánchez, A. G. and Sanchez, E. and Sanders, D. B. and Sapone, D. and Saponara, M. and Sarpa, E. and Sarron, F. and Sartori, S. and Sartoris, B. and Sassolas, B. and Sauniere, L. and Sauvage, M. and Sawicki, M. and Scaramella, R. and Scarlata, C. and Scharré, L. and Schaye, J. and Schewtschenko, J. A. and Schindler, J.-T. and Schinnerer, E. and Schirmer, M. and Schmidt, F. and Schmidt, F. and Schmidt, M. and Schneider, A. and Schneider, M. and Schneider, P. and Schöneberg, N. and Schrabback, T. and Schultheis, M. and Schulz, S. and Schuster, N. and Schwartz, J. and Sciotti, D. and Scodeggio, M. and Scognamiglio, D. and Scott, D. and Scottez, V. and Secroun, A. and Sefusatti, E. and Seidel, G. and Seiffert, M. and Sellentin, E. and Selwood, M. and Semboloni, E. and Sereno, M. and Serjeant, S. and Serrano, S. and Setnikar, G. and Shankar, F. and Sharples, R. M. and Short, A. and Shulevski, A. and Shuntov, M. and Sias, M. and Sikkema, G. and Silvestri, A. and Simon, P. and Sirignano, C. and Sirri, G. and Skottfelt, J. and Slezak, E. and Sluse, D. and Smith, G. P. and Smith, L. C. and Smith, R. E. and Smit, S. J. A. and Soldano, F. and Solheim, B. G. B. and Sorce, J. G. and Sorrenti, F. and Soubrie, E. and Spinoglio, L. and Spurio Mancini, A. and Stadel, J. and Stagnaro, L. and Stanco, L. and Stanford, S. A. and Starck, J.-L. and Stassi, P. and Steinwagner, J. and Stern, D. and Stone, C. and Strada, P. and Strafella, F. and Stramaccioni, D. and Surace, C. and Sureau, F. and Suyu, S. H. and Swindells, I. and Szafraniec, M. and Szapudi, I. and Taamoli, S. and Talia, M. and Tallada-Crespí, P. and Tanidis, K. and Tao, C. and Tarrío, P. and Tavagnacco, D. and Taylor, A. N. and Taylor, J. E. and Taylor, P. L. and Teixeira, E. M. and Tenti, M. and Teodoro Idiago, P. and Teplitz, H. I. and Tereno, I. and Tessore, N. and Testa, V. and Testera, G. and Tewes, M. and Teyssier, R. and Theret, N. and Thizy, C. and Thomas, P. D. and Toba, Y. and Toft, S. and Toledo-Moreo, R. and Tolstoy, E. and Tommasi, E. and Torbaniuk, O. and Torradeflot, F. and Tortora, C. and Tosi, S. and Tosti, S. and Trifoglio, M. and Troja, A. and Trombetti, T. and Tronconi, A. and Tsedrik, M. and Tsyganov, A. and Tucci, M. and Tutusaus, I. and Uhlemann, C. and Ulivi, L. and Urbano, M. and Vacher, L. and Vaillon, L. and Valageas, P. and Valdes, I. and Valentijn, E. A. and Valenziano, L. and Valieri, C. and Valiviita, J. and Van den Broeck, M. and Vassallo, T. and Vavrek, R. and Vega-Ferrero, J. and Venemans, B. and Venhola, A. and Ventura, S. and Verdoes Kleijn, G. and Vergani, D. and Verma, A. and Vernizzi, F. and Veropalumbo, A. and Verza, G. and Vescovi, C. and Vibert, D. and Viel, M. and Vielzeuf, P. and Viglione, C. and Viitanen, A. and Villaescusa-Navarro, F. and Vinciguerra, S. and Visticot, F. and Voggel, K. and von Wietersheim-Kramsta, M. and Vriend, W. J. and Wachter, S. and Walmsley, M. and Walth, G. and Walton, D. M. and Walton, N. A. and Wander, M. and Wang, L. and Wang, Y. and Weaver, J. R. and Weller, J. and Wetzstein, M. and Whalen, D. J. and Whittam, I. H. and Widmer, A. and Wiesmann, M. and Wilde, J. and Williams, O. R. and Winther, H.-A. and Wittje, A. and Wong, J. H. W. and Wright, A. H. and Yankelevich, V. and Yeung, H. W. and Yoon, M. and Youles, S. and Yung, L. Y. A. and Zacchei, A. and Zalesky, L. and Zamorani, G. and Zamorano Vitorelli, A. and Zanoni Marc, M. and Zennaro, M. and Zerbi, F. M. and Zinchenko, I. A. and Zoubian, J. and Zucca, E. and Zumalacarregui, M.},
   year={2025},
   month=apr, pages={A1} }

@article{Treu_2022,
doi = {10.3847/1538-4357/ac8158},
url = {https://doi.org/10.3847/1538-4357/ac8158},
year = {2022},
month = {aug},
publisher = {The American Astronomical Society},
volume = {935},
number = {2},
pages = {110},
author = {Treu, T. and Roberts-Borsani, G. and Bradac, M. and Brammer, G. and Fontana, A. and Henry, A. and Mason, C. and Morishita, T. and Pentericci, L. and Wang, X. and Acebron, A. and Bagley, M. and Bergamini, P. and Belfiori, D. and Bonchi, A. and Boyett, K. and Boutsia, K. and Calabró, A. and Caminha, G. B. and Castellano, M. and Dressler, A. and Glazebrook, K. and Grillo, C. and Jacobs, C. and Jones, T. and Kelly, P. L. and Leethochawalit, N. and Malkan, M. A. and Marchesini, D. and Mascia, S. and Mercurio, A. and Merlin, E. and Nanayakkara, T. and Nonino, M. and Paris, D. and Poggianti, B. and Rosati, P. and Santini, P. and Scarlata, C. and Shipley, H. V. and Strait, V. and Trenti, M. and Tubthong, C. and Vanzella, E. and Vulcani, B. and Yang, L.},
title = {The GLASS-JWST Early Release Science Program. I. Survey Design and Release Plans},
journal = {The Astrophysical Journal}
}

@article{Paris_2023,
doi = {10.3847/1538-4357/acda8a},
url = {https://doi.org/10.3847/1538-4357/acda8a},
year = {2023},
month = {jul},
publisher = {The American Astronomical Society},
volume = {952},
number = {1},
pages = {20},
author = {Paris, Diego and Merlin, Emiliano and Fontana, Adriano and Bonchi, Andrea and Brammer, Gabriel and Correnti, Matteo and Treu, Tommaso and Boyett, Kristan and Calabrò, Antonello and Castellano, Marco and Chen, Wenlei and Yang, Lilan and Glazebrook, Karl and Kelly, Patrick and Koekemoer, Anton M. and Leethochawalit, Nicha and Mascia, Sara and Mason, Charlotte and Morishita, Takahiro and Nonino, Mario and Pentericci, Laura and Polenta, Gianluca and Roberts-Borsani, Guido and Santini, Paola and Trenti, Michele and Vanzella, Eros and Vulcani, Benedetta and Windhorst, Rogier A. and Nanayakkara, Themiya and Wang, Xin},
title = {The GLASS-JWST Early Release Science Program. II. Stage I Release of NIRCam Imaging and Catalogs in the Abell 2744 Region},
journal = {The Astrophysical Journal}
}

@article{Bergamini_2023,
doi = {10.3847/1538-4357/acd643},
url = {https://doi.org/10.3847/1538-4357/acd643},
year = {2023},
month = {jul},
publisher = {The American Astronomical Society},
volume = {952},
number = {1},
pages = {84},
author = {Bergamini, Pietro and Acebron, Ana and Grillo, Claudio and Rosati, Piero and Caminha, Gabriel Bartosch and Mercurio, Amata and Vanzella, Eros and Mason, Charlotte and Treu, Tommaso and Angora, Giuseppe and Brammer, Gabriel B. and Meneghetti, Massimo and Nonino, Mario and Boyett, Kristan and Bradač, Maruša and Castellano, Marco and Fontana, Adriano and Morishita, Takahiro and Paris, Diego and Prieto-Lyon, Gonzalo and Roberts-Borsani, Guido and Roy, Namrata and Santini, Paola and Vulcani, Benedetta and Wang, Xin and Yang, Lilan},
title = {The GLASS-JWST Early Release Science Program. III. Strong-lensing Model of Abell 2744 and Its Infalling Regions},
journal = {The Astrophysical Journal}
}

@misc{Spergel_2015,
      title={Wide-Field InfrarRed Survey Telescope-Astrophysics Focused Telescope Assets WFIRST-AFTA 2015 Report}, 
      author={D. Spergel and N. Gehrels and C. Baltay and D. Bennett and J. Breckinridge and M. Donahue and A. Dressler and B. S. Gaudi and T. Greene and O. Guyon and C. Hirata and J. Kalirai and N. J. Kasdin and B. Macintosh and W. Moos and S. Perlmutter and M. Postman and B. Rauscher and J. Rhodes and Y. Wang and D. Weinberg and D. Benford and M. Hudson and W. -S. Jeong and Y. Mellier and W. Traub and T. Yamada and P. Capak and J. Colbert and D. Masters and M. Penny and D. Savransky and D. Stern and N. Zimmerman and R. Barry and L. Bartusek and K. Carpenter and E. Cheng and D. Content and F. Dekens and R. Demers and K. Grady and C. Jackson and G. Kuan and J. Kruk and M. Melton and B. Nemati and B. Parvin and I. Poberezhskiy and C. Peddie and J. Ruffa and J. K. Wallace and A. Whipple and E. Wollack and F. Zhao},
      year={2015},
      eprint={1503.03757},
      archivePrefix={arXiv},
      primaryClass={astro-ph.IM},
      url={https://arxiv.org/abs/1503.03757}, 
}

@misc{Akeson_2019,
      title={The Wide Field Infrared Survey Telescope: 100 Hubbles for the 2020s}, 
      author={Rachel Akeson and Lee Armus and Etienne Bachelet and Vanessa Bailey and Lisa Bartusek and Andrea Bellini and Dominic Benford and David Bennett and Aparna Bhattacharya and Ralph Bohlin and Martha Boyer and Valerio Bozza and Geoffrey Bryden and Sebastiano Calchi Novati and Kenneth Carpenter and Stefano Casertano and Ami Choi and David Content and Pratika Dayal and Alan Dressler and Olivier Doré and S. Michael Fall and Xiaohui Fan and Xiao Fang and Alexei Filippenko and Steven Finkelstein and Ryan Foley and Steven Furlanetto and Jason Kalirai and B. Scott Gaudi and Karoline Gilbert and Julien Girard and Kevin Grady and Jenny Greene and Puragra Guhathakurta and Chen Heinrich and Shoubaneh Hemmati and David Hendel and Calen Henderson and Thomas Henning and Christopher Hirata and Shirley Ho and Eric Huff and Anne Hutter and Rolf Jansen and Saurabh Jha and Samson Johnson and David Jones and Jeremy Kasdin and Patrick Kelly and Robert Kirshner and Anton Koekemoer and Jeffrey Kruk and Nikole Lewis and Bruce Macintosh and Piero Madau and Sangeeta Malhotra and Kaisey Mandel and Elena Massara and Daniel Masters and Julie McEnery and Kristen McQuinn and Peter Melchior and Mark Melton and Bertrand Mennesson and Molly Peeples and Matthew Penny and Saul Perlmutter and Alice Pisani and Andrés Plazas and Radek Poleski and Marc Postman and Clément Ranc and Bernard Rauscher and Armin Rest and Aki Roberge and Brant Robertson and Steven Rodney and James Rhoads and Jason Rhodes and Russell Ryan Jr. and Kailash Sahu and David Sand and Dan Scolnic and Anil Seth and Yossi Shvartzvald and Karelle Siellez and Arfon Smith and David Spergel and Keivan Stassun and Rachel Street and Louis-Gregory Strolger and Alexander Szalay and John Trauger and M. A. Troxel and Margaret Turnbull and Roeland van der Marel and Anja von der Linden and Yun Wang and David Weinberg and Benjamin Williams and Rogier Windhorst and Edward Wollack and Hao-Yi Wu and Jennifer Yee and Neil Zimmerman},
      year={2019},
      eprint={1902.05569},
      archivePrefix={arXiv},
      primaryClass={astro-ph.IM},
      url={https://arxiv.org/abs/1902.05569}, 
}

@misc{csstcollaboration2025introductionchinesespacestation,
      title={Introduction to the Chinese Space Station Survey Telescope (CSST)}, 
      author={CSST Collaboration and Yan Gong and Haitao Miao and Hu Zhan and Zhao-Yu Li and Jinyi Shangguan and Haining Li and Chao Liu and Xuefei Chen and Haibo Yuan and Jilin Zhou and Hui-Gen Liu and Cong Yu and Jianghui Ji and Zhaoxiang Qi and Jiacheng Liu and Zigao Dai and Xiaofeng Wang and Zhenya Zheng and Lei Hao and Jiangpei Dou and Yiping Ao and Zhenhui Lin and Kun Zhang and Wei Wang and Guotong Sun and Ran Li and Guoliang Li and Youhua Xu and Xinfeng Li and Shengyang Li and Peng Wu and Jiuxing Zhang and Bo Wang and Jinming Bai and Yi-Fu Cai and Zheng Cai and Jie Cao and Kwan Chuen Chan and Jin Chang and Xiaodian Chen and Xuelei Chen and Yuqin Chen and Yun Chen and Wei Cui and Subo Dong and Pu Du and Wenying Duan and Junhui Fan and LuLu Fan and Zhou Fan and Zuhui Fan and Taotao Fang and Jianning Fu and Liping Fu and Zhensen Fu and Jian Gao and Shenghong Gu and Yidong Gu and Qi Guo and Zhanwen Han and Bin Hu and Zhiqi Huang and Luis C. Ho and Linhua Jiang and Ning Jiang and Yipeng Jing and Xi Kang and Xu Kong and Cheng Li and Chengyuan Li and Di Li and Jing Li and Nan Li and Yang A. Li and Shilong Liao and Weipeng Lin and Fengshan Liu and Jifeng Liu and Xiangkun Liu and Zhuokai Liu and Ruiqing Mao and Shude Mao and Xianmin Meng and Xiaoying Pang and Xiyan Peng and Yingjie Peng and Huanyuan Shan and Juntai Shen and Shiyin Shen and Zhiqiang Shen and Sheng-Cai Shi and Yong Shi and Siyuan Tan and Hao Tian and Jianmin Wang and Jun-Xian Wang and Xin Wang and Yuting Wang and Hong Wu and Jingwen Wu and Xuebing Wu and Chun Xu and Xiang-Xiang Xue and Yongquan Xue and Ji Yang and Xiaohu Yang and Qijun Yao and Fangting Yuan and Zhen Yuan and Jun Zhang and Pengjie Zhang and Tianmeng Zhang and Wei Zhang and Xin Zhang and Gang Zhao and Gongbo Zhao and Hongen Zhong and Jing Zhong and Liyong Zhou and Wei Zhu and Ying Zu},
      year={2025},
      eprint={2507.04618},
      archivePrefix={arXiv},
      primaryClass={astro-ph.IM},
      url={https://arxiv.org/abs/2507.04618}, 
}

@article{Zhan_2021,
    author = {Zhan, H.},
    title = {The wide-field multiband imaging and slitless spectroscopy survey to be carried out by the Survey Space Telescope of China Manned Space Program.},
    doi = {10.1360/TB-2021-0016},
    journal = {Kexue Tongbao/Chinese Science Bulletin},
    volume = {66},
    number = {11},
    pages = {1290–1298},
    year = {2021}
}

@misc{Nie_2025,
      title={Toward High-Precision Astrometry with CSST Using Multi-Gaussian Fitting of PSF}, 
      author={Jialu Nie and Peng Wei and Zihuang Cao and Yibo Yan and Chao Liu and Hao Tian and Xin Zhang and Haijun Tian},
      year={2025},
      eprint={2508.10329},
      archivePrefix={arXiv},
      primaryClass={astro-ph.IM},
      url={https://arxiv.org/abs/2508.10329}, 
}

@article{Gong_2019,
   title={Cosmology from the Chinese Space Station Optical Survey (CSS-OS)},
   volume={883},
   ISSN={1538-4357},
   url={http://dx.doi.org/10.3847/1538-4357/ab391e},
   DOI={10.3847/1538-4357/ab391e},
   number={2},
   journal={The Astrophysical Journal},
   publisher={American Astronomical Society},
   author={Gong, Yan and Liu, Xiangkun and Cao, Ye and Chen, Xuelei and Fan, Zuhui and Li, Ran and Li, Xiao-Dong and Li, Zhigang and Zhang, Xin and Zhan, Hu},
   year={2019},
   month=oct, pages={203} }

@article{Gong_2025,
    author = "Gong, Yan and others",
    title = "{Future cosmology: New physics and opportunity from the China Space Station Telescope (CSST)}",
    eprint = "2501.15023",
    archivePrefix = "arXiv",
    primaryClass = "astro-ph.CO",
    doi = "10.1007/s11433-025-2646-2",
    journal = "Sci. China Phys. Mech. Astron.",
    volume = "68",
    number = "8",
    pages = "280402",
    year = "2025"
}

@article{Cao_2018,
   title={Testing photometric redshift measurements with filter definition of the Chinese Space Station Optical Survey (CSS-OS)},
   ISSN={1365-2966},
   url={http://dx.doi.org/10.1093/mnras/sty1980},
   DOI={10.1093/mnras/sty1980},
   journal={Monthly Notices of the Royal Astronomical Society},
   publisher={Oxford University Press (OUP)},
   author={Cao, Ye and Gong, Yan and Meng, Xian-Min and Xu, Cong K and Chen, Xuelei and Guo, Qi and Li, Ran and Liu, Dezi and Xue, Yongquan and Cao, Li and Fu, Xiyang and Zhang, Xin and Wang, Shen and Zhan, Hu},
   year={2018},
   month=jul 
}

@article{Bartelmann_2001,
   title={Weak gravitational lensing},
   volume={340},
   ISSN={0370-1573},
   url={http://dx.doi.org/10.1016/S0370-1573(00)00082-X},
   DOI={10.1016/s0370-1573(00)00082-x},
   number={4–5},
   journal={Physics Reports},
   publisher={Elsevier BV},
   author={Bartelmann, Matthias and Schneider, Peter},
   year={2001},
   month=jan, pages={291–472} }

@article{Kaiser_1995,
   title={A Method for Weak Lensing Observations},
   volume={449},
   ISSN={1538-4357},
   url={http://dx.doi.org/10.1086/176071},
   DOI={10.1086/176071},
   journal={The Astrophysical Journal},
   publisher={American Astronomical Society},
   author={Kaiser, Nick and Squires, Gordon and Broadhurst, Tom},
   year={1995},
   month=aug, pages={460} }

@INPROCEEDINGS{Krist1993,
       author = {{Krist}, J.},
        title = "{Tiny Tim : an HST PSF Simulator}",
    booktitle = {Astronomical Data Analysis Software and Systems II},
         year = 1993,
       editor = {{Hanisch}, R.~J. and {Brissenden}, R.~J.~V. and {Barnes}, J.},
       series = {Astronomical Society of the Pacific Conference Series},
       volume = {52},
        month = jan,
        pages = {536},
       adsurl = {https://ui.adsabs.harvard.edu/abs/1993ASPC...52..536K}
}

@article{Anderson_2000,
doi = {10.1086/316632},
url = {https://dx.doi.org/10.1086/316632},
year = {2000},
month = {oct},
publisher = {The University of Chicago Press},
volume = {112},
number = {776},
pages = {1360},
author = {Anderson, Jay and King, Ivan R.},
title = {Toward High‐Precision Astrometry with WFPC2. I. Deriving an Accurate Point‐Spread Function},
journal = {Publications of the Astronomical Society of the Pacific}
}

@MISC{Anderson2006,
       author = {{Anderson}, Jay and {King}, Ivan R.},
        title = "{PSFs, Photometry, and Astronomy for the ACS/WFC}",
 howpublished = {Instrument Science Report ACS 2006-01, 34 pages},
         year = 2006,
        month = feb,
        pages = {1},
       adsurl = {https://ui.adsabs.harvard.edu/abs/2006acs..rept....1A}
}

@article{Ngolè_2016,
doi = {10.1088/0266-5611/32/12/124001},
url = {https://dx.doi.org/10.1088/0266-5611/32/12/124001},
year = {2016},
month = {nov},
publisher = {IOP Publishing},
volume = {32},
number = {12},
pages = {124001},
author = {Ngolè, F and Starck, J-L and Okumura, K and Amiaux, J and Hudelot, P},
title = {Constraint matrix factorization for space variant PSFs field restoration},
journal = {Inverse Problems}
}

@INPROCEEDINGS{Bertin_2011,
       author = {{Bertin}, E.},
        title = "{Automated Morphometry with SExtractor and PSFEx}",
    booktitle = {Astronomical Data Analysis Software and Systems XX},
         year = 2011,
       editor = {{Evans}, I.~N. and {Accomazzi}, A. and {Mink}, D.~J. and {Rots}, A.~H.},
       series = {Astronomical Society of the Pacific Conference Series},
       volume = {442},
        month = jul,
        pages = {435},
       adsurl = {https://ui.adsabs.harvard.edu/abs/2011ASPC..442..435B}
}

@article{Liaudat_2023,
   title={Rethinking data-driven point spread function modeling with a differentiable optical model},
   volume={39},
   ISSN={1361-6420},
   url={http://dx.doi.org/10.1088/1361-6420/acb664},
   DOI={10.1088/1361-6420/acb664},
   number={3},
   journal={Inverse Problems},
   publisher={IOP Publishing},
   author={Liaudat, Tobias and Starck, Jean-Luc and Kilbinger, Martin and Frugier, Pierre-Antoine},
   year={2023},
   month=feb, pages={035008} }

@article{Jia_2021,
   title={Point spread function estimation for wide field small aperture telescopes with deep neural networks and calibration data},
   volume={505},
   ISSN={1365-2966},
   url={http://dx.doi.org/10.1093/mnras/stab1461},
   DOI={10.1093/mnras/stab1461},
   number={4},
   journal={Monthly Notices of the Royal Astronomical Society},
   publisher={Oxford University Press (OUP)},
   author={Jia, Peng and Wu, Xuebo and Li, Zhengyang and Li, Bo and Wang, Weihua and Liu, Qiang and Popowicz, Adam and Cai, Dongmei},
   year={2021},
   month=may, pages={4717–4725} }

@article{Jia_2020,
doi = {10.3847/1538-3881/ab7b79},
url = {https://dx.doi.org/10.3847/1538-3881/ab7b79},
year = {2020},
month = {apr},
publisher = {The American Astronomical Society},
volume = {159},
number = {4},
pages = {183},
author = {Jia, Peng and Wu, Xuebo and Yi, Huang and Cai, Bojun and Cai, Dongmei},
title = {PSF–NET: A Nonparametric Point-spread Function Model for Ground-based Optical Telescopes},
journal = {The Astronomical Journal}
}

@article{Lanusse_2021,
    author = {Lanusse, François and Mandelbaum, Rachel and Ravanbakhsh, Siamak and Li, Chun-Liang and Freeman, Peter and Póczos, Barnabás},
    title = {Deep generative models for galaxy image simulations},
    journal = {Monthly Notices of the Royal Astronomical Society},
    volume = {504},
    number = {4},
    pages = {5543-5555},
    year = {2021},
    month = {05},
    issn = {0035-8711},
    doi = {10.1093/mnras/stab1214},
    url = {https://doi.org/10.1093/mnras/stab1214},
    eprint = {https://academic.oup.com/mnras/article-pdf/504/4/5543/38036124/stab1214.pdf},
}

@inproceedings{Carney_2024,
author = {Jonathan Carney and Hank Corbett and William Marshall and Nicholas Law and Shannon Fitton and Ramses Gonzalez and Lawrence Machia and Thomas Proctor and Alan Vasquez},
title = {PSF modeling with deep learning for the Argus Optical Array},
volume = {13101},
booktitle = {Software and Cyberinfrastructure for Astronomy VIII},
editor = {Jorge Ibsen and Gianluca Chiozzi},
organization = {International Society for Optics and Photonics},
publisher = {SPIE},
pages = {131010O},
year = {2024},
doi = {10.1117/12.3020465},
URL = {https://doi.org/10.1117/12.3020465}
}

@misc{Ni_2024,
      title={PI-AstroDeconv: A Physics-Informed Unsupervised Learning Method for Astronomical Image Deconvolution}, 
      author={Shulei Ni and Yisheng Qiu and Yunchun Chen and Zihao Song and Hao Chen and Xuejian Jiang and Huaxi Chen},
      year={2024},
      eprint={2403.01692},
      archivePrefix={arXiv},
      primaryClass={astro-ph.IM},
      url={https://arxiv.org/abs/2403.01692}, 
}

@misc{Bai_2025,
      title={An Ultra-Fast Image Simulation Technique with Spatially Variable Point Spread Functions}, 
      author={Zeyu Bai and Peng Jia and Jiameng Lv and Xiang Zhang and Wennan Xiang and Lin Nie},
      year={2025},
      eprint={2502.10015},
      archivePrefix={arXiv},
      primaryClass={astro-ph.IM},
      url={https://arxiv.org/abs/2502.10015}, 
}

@ARTICLE{2015A&C....10..121R,
       author = {{Rowe}, B.~T.~P. and {Jarvis}, M. and {Mandelbaum}, R. and {Bernstein}, G.~M. and {Bosch}, J. and {Simet}, M. and {Meyers}, J.~E. and {Kacprzak}, T. and {Nakajima}, R. and {Zuntz}, J. and {Miyatake}, H. and {Dietrich}, J.~P. and {Armstrong}, R. and {Melchior}, P. and {Gill}, M.~S.~S.},
        title = "{GALSIM: The modular galaxy image simulation toolkit}",
      journal = {Astronomy and Computing},
         year = 2015,
        month = apr,
       volume = {10},
        pages = {121-150},
          doi = {10.1016/j.ascom.2015.02.002},
archivePrefix = {arXiv},
       eprint = {1407.7676},
 primaryClass = {astro-ph.IM},
       adsurl = {https://ui.adsabs.harvard.edu/abs/2015A&C....10..121R}
}

@misc{Ban_2025,
      title={Mock Observations for the CSST Mission: End-to-End Performance Modeling of Optical System}, 
      author={Zhang Ban and Xiao-Bo Li and Xun Yang and Yu-Xi Jiang and Hong-Cai Ma and Wei Wang and Jin-guang Lv and Cheng-Liang Wei and De-Zi Liu and Guo-Liang Li and Chao Liu and Nan Li and Ran Li and Peng Wei},
      year={2025},
      eprint={2511.06936},
      archivePrefix={arXiv},
      primaryClass={astro-ph.IM},
      url={https://arxiv.org/abs/2511.06936}, 
}

@article{Ban_2022,
    year = {2022},
    author = {Ban, Zhang and Li, Xiaobo and Yang, Xun and Jiang, Yuxi},
    title = {Analysis and Research on End-to-end Optical Image Quality of Large Aperture Off-axis Space Telescope},
    journal = {Journal of System Simulation},
    volume={34},
    doi={https://doi.org/10.16182/j.issn1004731x.joss.21-0582}
    }

@misc{ling2026,
      title={Selecting Optimal Stellar Calibration Fields for the CSST Imaging Survey}, 
      author={Chenxiaoji Ling and Juanjuan Ren and Li Shao and Zhimin Zhou and Peng Wei and Youhua Xu and Jinyu Hu and Xin Zhang and Su Yao and Hu Zhan and Chao Liu},
      year={2026},
      eprint={2602.11864},
      archivePrefix={arXiv},
      primaryClass={astro-ph.SR},
      url={https://arxiv.org/abs/2602.11864}, 
}

@ARTICLE{Bertin_1996,
       author = {{Bertin}, E. and {Arnouts}, S.},
        title = "{SExtractor: Software for source extraction.}",
      journal = {\aaps},
         year = 1996,
        month = jun,
       volume = {117},
        pages = {393-404},
          doi = {10.1051/aas:1996164},
       adsurl = {https://ui.adsabs.harvard.edu/abs/1996A\&AS..117..393B}
}

@article{Wang_2004,
  author={Zhou Wang and Bovik, A.C. and Sheikh, H.R. and Simoncelli, E.P.},
  journal={IEEE Transactions on Image Processing}, 
  title={Image quality assessment: from error visibility to structural similarity}, 
  year={2004},
  volume={13},
  number={4},
  pages={600-612},
  doi={10.1109/TIP.2003.819861}}

@article{Avanaki_2009,
  author={Avanaki, A.N.},
  journal={OPTICAL REVIEW}, 
  title={Exact global histogram specification optimized for structural similarity}, 
  year={2009},
  volume={16},
  pages={613–621},
  doi={https://doi.org/10.1007/s10043-009-0119-z}}

@article{Paulin_Henriksson_2008,
   title={Point spread function calibration requirements 
for dark energy from cosmic shear},
   volume={484},
   ISSN={1432-0746},
   url={http://dx.doi.org/10.1051/0004-6361:20079150},
   DOI={10.1051/0004-6361:20079150},
   number={1},
   journal={Astronomy \&amp; Astrophysics},
   publisher={EDP Sciences},
   author={Paulin-Henriksson, S. and Amara, A. and Voigt, L. and Refregier, A. and Bridle, S. L.},
   year={2008},
   month=mar, pages={67–77} }

@article{Paulin_Henriksson_2009,
   title={Optimal point spread function modeling for weak lensing: 
complexity and sparsity},
   volume={500},
   ISSN={1432-0746},
   url={http://dx.doi.org/10.1051/0004-6361/200811061},
   DOI={10.1051/0004-6361/200811061},
   number={2},
   journal={Astronomy \&amp; Astrophysics},
   publisher={EDP Sciences},
   author={Paulin-Henriksson, S. and Refregier, A. and Amara, A.},
   year={2009},
   month=apr, pages={647–655} }

@article{Hirata_2003,
    author = {Hirata, Christopher and Seljak, Uroš},
    title = {Shear calibration biases in weak-lensing surveys},
    journal = {Monthly Notices of the Royal Astronomical Society},
    volume = {343},
    number = {2},
    pages = {459-480},
    year = {2003},
    month = {08},
    issn = {0035-8711},
    doi = {10.1046/j.1365-8711.2003.06683.x},
    url = {https://doi.org/10.1046/j.1365-8711.2003.06683.x},
    eprint = {https://academic.oup.com/mnras/article-pdf/343/2/459/18417667/343-2-459.pdf},
}

@article{Heymans_2005,
    author = {Heymans, Catherine and Brown, Michael L. and Barden, Marco and Caldwell, John A. R. and Jahnke, Knud and Peng, Chien Y. and Rix, Hans-Walter and Taylor, Andy and Beckwith, Steven V. W. and Bell, Eric F. and Borch, Andrea and Häußler, Boris and Jogee, Shardha and McIntosh, Daniel H. and Meisenheimer, Klaus and Sánchez, Sebastian F. and Somerville, Rachel and Wisotzki, Lutz and Wolf, Christian},
    title = {Cosmological weak lensing with the HST GEMS survey},
    journal = {Monthly Notices of the Royal Astronomical Society},
    volume = {361},
    number = {1},
    pages = {160-176},
    year = {2005},
    month = {07},
    issn = {0035-8711},
    doi = {10.1111/j.1365-2966.2005.09152.x},
    url = {https://doi.org/10.1111/j.1365-2966.2005.09152.x},
    eprint = {https://academic.oup.com/mnras/article-pdf/361/1/160/18654533/361-1-160.pdf},
}

@article{Mandelbaum_2017,
    author = {Mandelbaum, Rachel and Miyatake, Hironao and Hamana, Takashi and Oguri, Masamune and Simet, Melanie and Armstrong, Robert and Bosch, James and Murata, Ryoma and Lanusse, François and Leauthaud, Alexie and Coupon, Jean and More, Surhud and Takada, Masahiro and Miyazaki, Satoshi and Speagle, Joshua S and Shirasaki, Masato and Sifón, Cristóbal and Huang, Song and Nishizawa, Atsushi J and Medezinski, Elinor and Okura, Yuki and Okabe, Nobuhiro and Czakon, Nicole and Takahashi, Ryuichi and Coulton, William R and Hikage, Chiaki and Komiyama, Yutaka and Lupton, Robert H and Strauss, Michael A and Tanaka, Masayuki and Utsumi, Yousuke},
    title = {The first-year shear catalog of the Subaru Hyper Suprime-Cam Subaru Strategic Program Survey},
    journal = {Publications of the Astronomical Society of Japan},
    volume = {70},
    number = {SP1},
    pages = {S25},
    year = {2017},
    month = {12},
    issn = {0004-6264},
    doi = {10.1093/pasj/psx130},
    url = {https://doi.org/10.1093/pasj/psx130},
    eprint = {https://academic.oup.com/pasj/article-pdf/70/SP1/S25/54675906/pasj_70_sp1_s25.pdf},
}

@article{Guinot_2022,
   title={ShapePipe: A new shape measurement pipeline and weak-lensing application to UNIONS/CFIS data},
   volume={666},
   ISSN={1432-0746},
   url={http://dx.doi.org/10.1051/0004-6361/202141847},
   DOI={10.1051/0004-6361/202141847},
   journal={Astronomy \&amp; Astrophysics},
   publisher={EDP Sciences},
   author={Guinot, Axel and Kilbinger, Martin and Farrens, Samuel and Peel, Austin and Pujol, Arnau and Schmitz, Morgan and Starck, Jean-Luc and Erben, Thomas and Gavazzi, Raphael and Gwyn, Stephen and Hudson, Michael J. and Hildebrandt, Hendrik and Tobias, Liaudat and Miller, Lance and Spitzer, Isaac and Van Waerbeke, Ludovic and Cuillandre, Jean-Charles and Fabbro, Sébastien and McConnachie, Alan and Mellier, Yannick},
   year={2022},
   month=oct, pages={A162} }

@article{Zuntz_2018,
   title={Dark Energy Survey Year 1 results: weak lensing shape catalogues},
   volume={481},
   ISSN={1365-2966},
   url={http://dx.doi.org/10.1093/mnras/sty2219},
   DOI={10.1093/mnras/sty2219},
   number={1},
   journal={Monthly Notices of the Royal Astronomical Society},
   publisher={Oxford University Press (OUP)},
   author={Zuntz, J and Sheldon, E and Samuroff, S and Troxel, M A and Jarvis, M and MacCrann, N and Gruen, D and Prat, J and Sánchez, C and Choi, A and Bridle, S L and Bernstein, G M and Dodelson, S and Drlica-Wagner, A and Fang, Y and Gruendl, R A and Hoyle, B and Huff, E M and Jain, B and Kirk, D and Kacprzak, T and Krawiec, C and Plazas, A A and Rollins, R P and Rykoff, E S and Sevilla-Noarbe, I and Soergel, B and Varga, T N and Abbott, T M C and Abdalla, F B and Allam, S and Annis, J and Bechtol, K and Benoit-Lévy, A and Bertin, E and Buckley-Geer, E and Burke, D L and Carnero Rosell, A and Kind, M Carrasco and Carretero, J and Castander, F J and Crocce, M and Cunha, C E and D’Andrea, C B and da Costa, L N and Davis, C and Desai, S and Diehl, H T and Dietrich, J P and Doel, P and Eifler, T F and Estrada, J and Evrard, A E and Neto, A Fausti and Fernandez, E and Flaugher, B and Fosalba, P and Frieman, J and García-Bellido, J and Gaztanaga, E and Gerdes, D W and Giannantonio, T and Gschwend, J and Gutierrez, G and Hartley, W G and Honscheid, K and James, D J and Jeltema, T and Johnson, M W G and Johnson, M D and Kuehn, K and Kuhlmann, S and Kuropatkin, N and Lahav, O and Li, T S and Lima, M and Maia, M A G and March, M and Martini, P and Melchior, P and Menanteau, F and Miller, C J and Miquel, R and Mohr, J J and Neilsen, E and Nichol, R C and Ogando, R L C and Roe, N and Romer, A K and Roodman, A and Sanchez, E and Scarpine, V and Schindler, R and Schubnell, M and Smith, M and Smith, R C and Soares-Santos, M and Sobreira, F and Suchyta, E and Swanson, M E C and Tarle, G and Thomas, D and Tucker, D L and Vikram, V and Walker, A R and Wechsler, R H and Zhang, Y},
   year={2018},
   month=aug, pages={1149–1182} }

@misc{wei_2025,
      title={Mock Observations for the CSST Mission: Main Surveys--An Overview of Framework and Simulation Suite}, 
      author={Cheng-Liang Wei and Guo-Liang Li and Yue-Dong Fang and Xin Zhang and Yu Luo and Hao Tian and De-Zi Liu and Xian-Ming Meng and Zhang Ban and Xiao-Bo Li and Zun Luo and Jing-Tian Xian and Wei Wang and Xi-Yan Peng and Nan Li and Ran Li and Li Shao and Tian-Meng Zhang and Jing Tang and Yang Chen and Zhao-Xiang Qi and Zi-Huang Cao and Huan-Yuan Shan and Lin Nie and Lei Wang and Zizhao He and Rui-Biao Luo and Quan-Yu Liu and Zhao-Jun Yan},
      year={2025},
      eprint={2511.06970},
      archivePrefix={arXiv},
      primaryClass={astro-ph.IM},
      url={https://arxiv.org/abs/2511.06970}, 
}

@ARTICLE{Bai2024,
       author = {{Bai}, Weimin and {Wang}, Yifei and {Chen}, Wenzheng and {Sun}, He},
        title = "{An Expectation-Maximization Algorithm for Training Clean Diffusion Models from Corrupted Observations}",
      journal = {arXiv e-prints},
         year = 2024,
        month = jul,
          eid = {arXiv:2407.01014},
        pages = {arXiv:2407.01014},
          doi = {10.48550/arXiv.2407.01014},
archivePrefix = {arXiv},
       eprint = {2407.01014},
 primaryClass = {cs.CV},
       adsurl = {https://ui.adsabs.harvard.edu/abs/2024arXiv240701014B}
}

@ARTICLE{Barco2025,
       author = {{Barco}, Gabriel Missael and {Adam}, Alexandre and {Stone}, Connor and {Hezaveh}, Yashar and {Perreault-Levasseur}, Laurence},
        title = "{Tackling the Problem of Distributional Shifts: Correcting Misspecified, High-dimensional Data-driven Priors for Inverse Problems}",
      journal = {\apj},
         year = 2025,
        month = feb,
       volume = {980},
       number = {1},
          eid = {108},
        pages = {108},
          doi = {10.3847/1538-4357/ad9b92},
archivePrefix = {arXiv},
       eprint = {2407.17667},
 primaryClass = {astro-ph.IM},
       adsurl = {https://ui.adsabs.harvard.edu/abs/2025ApJ...980..108B}
}

@ARTICLE{Hosseintabar2025,
       author = {{Hosseintabar}, Danial and {Chen}, Fan and {Daras}, Giannis and {Torralba}, Antonio and {Daskalakis}, Constantinos},
        title = "{DiffEM: Learning from Corrupted Data with Diffusion Models via Expectation Maximization}",
      journal = {arXiv e-prints},
         year = 2025,
        month = oct,
          eid = {arXiv:2510.12691},
        pages = {arXiv:2510.12691},
          doi = {10.48550/arXiv.2510.12691},
archivePrefix = {arXiv},
       eprint = {2510.12691},
 primaryClass = {stat.ML},
       adsurl = {https://ui.adsabs.harvard.edu/abs/2025arXiv251012691H}
}
\bibliographystyle{aasjournal}



\end{document}